\documentclass[a4paper,11pt]{article}
\pdfoutput=1 

\usepackage{jcappub} 

\usepackage[T1]{fontenc} 
\usepackage{appendix}
\usepackage{subcaption}
\usepackage{hyperref}

\hypersetup{
    colorlinks=true,
    urlcolor=blue,
    }

\usepackage{graphicx}
\usepackage{amssymb}
\usepackage{color}
\usepackage{tensor}

\newenvironment{mytabular}[1][1]{
  \renewcommand*{\arraystretch}{#1}
  \tabular
}{
  \endtabular
}

\title{Constraining Post-Newtonian Parameters with the Cosmic Microwave Background}

\author[a,b]{Daniel B. Thomas,}
\author [b]{Theodore Anton,}
\author[b]{Timothy Clifton}
\author[a,c]{ and Philip Bull}

\affiliation[a]{Jodrell Bank Centre for Astrophysics, The University of Manchester, UK}
\affiliation[b]{Department of Physics and Astronomy, Queen Mary University of London, UK}
\affiliation[c]{Department of Physics and Astronomy, University of Western Cape, Cape Town, South Africa}

\emailAdd{dan.b.thomas1@gmail.com}
\emailAdd{t.j.anton@qmul.ac.uk}
\emailAdd{t.clifton@qmul.ac.uk}
\emailAdd{phil.bull@manchester.ac.uk}

\abstract{
The Parameterised Post-Newtonian (PPN) approach is the default framework for performing precision tests of gravity in nearby astrophysical systems. In recent works we have extended this approach for cosmological applications, and in this paper we use observations of the anisotropies in the Cosmic Microwave Background to constrain the time variation of the PPN parameters $\alpha$ and $\gamma$ between last scattering and the present day. We find their time-averages over cosmological history should be within $\sim 20\%$ of their values in GR, with $\bar{\alpha}=0.89^{+0.08}_{-0.09}$ and $\bar{\gamma}=0.90^{+0.07}_{-0.08}$ at the $68\%$ confidence level. We also constrain the time derivatives of these parameters, and find that their present-day values should be within a factor of two of the best Solar System constraints. Many of these results have no counter-part from Solar System observations, and are entirely new constraints on the gravitational interaction. In all cases, we find that the data strongly prefer $\bar{\alpha}\simeq \bar{\gamma}$, meaning that observers would typically find local gravitational physics to be compatible with GR, despite considerable variation of $\alpha$ and $\gamma$ being allowed over cosmic history. This study lays the groundwork for future precision tests of gravity that combine observations made over all cosmological and astrophysical scales of length and time.
}

\begin{document}
\maketitle
\flushbottom

\newpage

\section{Introduction}

Einstein's theory of General Relativity (GR) provides us with a rich variety of `post-Newtonian' gravitational potentials, with sources that correspond to a wide array of different types of energy density. These include the gravitational fields of kinetic and internal energies, as well as those of lower-order gravitational potential energies themselves, which arise as a direct result of the non-linearity of the field equations of the theory. Investigating the gravitational phenomenology that result from this zoo of potentialities has been one of the key goals of gravitational physics over the past century, and we are now at a stage where almost all of them are very tightly constrained by observations made within the Solar System and other nearby astrophysical systems \cite{Will_2014}.

Part of the reason for the great success of this field, other than the advancement of the required technologies, is the development of a framework that clearly links the possible gravitational potentials to observational phenomena: the Parameterised Post-Newtonian (PPN) framework \cite{Will_1993}. The philosophy behind this approach is to constrain the coupling constants between the gravitational potentials and the sources of those gravitational potentials (mass, kinetic energy, gravitational potential energy, {\it etc}). With a minimal number of assumptions about the nature of gravity, it is then possible to use observations to constrain the amplitude of any given type of gravitational field, and hence to restrict the allowed values of the coupling constants that control them. This approach has allowed gravity to be constrained {\it without} specifying individual candidate theories, which would be an endless task.

In a series of recent works, we have extended this approach for use in cosmology, in a framework dubbed `Parameterised Post-Newtonian Cosmology' (PPNC) \cite{Sanghai_2017, Sanghai_2019, Anton_2021, Thomas_2023, Clifton_2024}. Our primary goal in doing this has been to maintain the theory-independence of the approach, while making minimal adjustments to allow for consistent cosmological models to be constructed. The end product is a cosmological space-time that is isometric to the geometry used in the PPN approach in any region of space smaller than the homogeneity scale ($\sim 100 \, {\rm Mpc}$), but which is globally an expanding Friedmann-Robertson-Walker (FRW) universe with perturbations. The required agreement between our cosmological construction, and the small-scale geometry used in the PPN approach, means that we are led to a set of Friedmann and cosmological perturbation equations that are specified directly in terms of the PPN parameters.

We expect our PPNC approach to be useful for gaining theory-independent constraints on gravity from cosmological observations, in a directly analogous way to the constraints that have been achieved using the PPN approach to gravity in nearby astrophysical systems. Indeed, it is exactly the same set of parameters that occur in both formalisms (with some minimal adjustments made for consistency in the cosmological context). The PPNC formalism therefore allows tests of gravity to be extended out of the Solar System and the Milky Way, and into the Universe at large. In particular, it allows for constraints to be imposed on the time-dependence of the PPN parameters as the Universe expands, in a direct extension of the idea of a varying Newton's constant \cite{uzan2003fundamental}. It also allows tests of gravity to be performed in wildly different physical environments to anything we find in the nearby Universe.

In this work we use observations of the anisotropies in the Cosmic Microwave Background (CMB), in order to constrain the degrees of freedom of the PPNC formalism, and hence to constrain the time evolution of the PPN parameters all the way from the time of last scattering through to the present day. This is a first step in using the PPNC formalism to interpret a complicated cosmological data set, but nonetheless allows us to find interesting constraints. We consider this work to lay the foundation for more comprehensive studies with other cosmological data sets, and combinations of data sets, resulting in the strongest possible constraints and mitigating the need for assumptions.

We consider our approach to be complementary to other attempts to parameterize alternative theories of gravity in cosmology (see e.g. \cite{1995ApJ...455....7M, Hu_2007, amin2008subhorizon, Skordis_2009, Baker_2011, Baker_2013, Bellini_2014, Clifton_2012, Ishak_2018} for reviews), as well as the calculation of the CMB anisotropies in those theories (see e.g. \cite{Brax_2012, refId0, PhysRevD.97.023520, PhysRevD.100.063509, Frusciante_2020, Joudaki_2022}). Our goal is to consider minimal deviations from the general-relativistic $\Lambda$CDM model, in order to use precise cosmological observables to place precision constraints on possible deviations from that theory, in the same way that the PPN formalism is used to place tight constraints on theories that deviate minimally from General Relativity in the Solar System. This complements other approaches that are designed to accommodate completely general behaviour in the scalar sector of cosmological perturbation theory \cite{1995ApJ...455....7M, Hu_2007, amin2008subhorizon}, or to exhaustively study all possible behaviours within classes of theories with specified field content \cite{Baker_2011, Baker_2013, Bellini_2014}. Our approach is conservative, in that it does not introduce anything beyond the most minimal possible deviations from General Relativity while making no assumptions about any additional field content of the underlying theory. Our approach is also comprehensive, in that it allows for both linear and non-linear sub-horizon perturbations, as well as large-scale background and super-horizon perturbations, to all be consistently described by the same parameters. The PPNC approach is also immediately compatible with the PPN approach in the appropriate limit, making it very simple and easy to combine constraints from disparate sets of observables, and thereby ensuring that cosmological behaviour is consistent with Solar System constraints (without, for example, needing to invoke an unseen screening mechanism).

The plan of this paper is as follows: in Section 2 we give a concise presentation of the PPNC framework, as well as how we implemented it into a Boltzmann code, and the MCMC approach used to infer parameter constraints. We then give our results in Section 3, before concluding in Section 4.

\section{Theory}
\label{theory}

In this Section we will outline the essential features of the PPNC construction, the way we implemented this approach into the Cosmic Linear Anisotropy Solving System (CLASS) \cite{lesgourgues2011cosmic}, and the methods used to obtain the constraints on the parameters.

\subsection{Parameterised Post-Newtonian Cosmology}

The PPNC formalism is constructed by gluing together numerous regions of space-time, each of which is internally described by the PPN test metric from Ref. \cite{Will_1993}. Even though each individual region of space-time is being described by a post-Newtonian expansion about Minkowski space, the construction that results from this process is globally a cosmology, which geometrically is close to homogeneous and isotropic at all points, and which is capable of describing scales of length and time that are appropriate for an expanding Universe \cite{Sanghai_2017}. 

While the small-scale limit of perturbations about this geometry are automatically taken care of by the post-Newtonian fields of the original test metric, the large-scale limits can be derived from a separate Universe approach \cite{Sanghai_2019}, and an interpolation between large- and small-scale limits of the perturbation equations can be performed \cite{Thomas_2023}. The results of this can then be complemented by a momentum constraint equation, which can be derived from a combination of considering vector potentials on small scales and by boosting observers on large scales \cite{Anton_2021}.

The end result of this work is the following description of a spatially-flat cosmology:
\begin{equation} 
\mathrm{d}s^2=a(\tau)^2\left[-(1-2{\Phi})\mathrm{d}{\tau}^2+(1+2{\Psi})\delta_{ij}\mathrm{d}{x}^i\mathrm{d}{x}^j \right] \label{eqn_flrw}\text{,}
\end{equation}
where $a(\tau)$ is the scale factor as a function of conformal time, and $\Phi$ and $\Psi$ are scalar perturbations in longitudinal gauge.
The Friedmann equations obeyed by the scale factor take the form
\begin{eqnarray}
\mathcal{H}^2 &=& \frac{8 \pi G a^2}{3}\, \gamma \, \bar{\rho}-\frac{2a^2}{3} \, \gamma_c \label{eqn_ppncbkgd1}\\
\mathcal{H}_{,\tau} &=& -\frac{4 \pi G a^2}{3} \, \alpha \, \bar{\rho}+\frac{a^2}{3} \, \alpha_c\, \text{,}\label{eqn_ppncbkgd2}
\end{eqnarray}
where $\mathcal{H} \equiv a_{,\tau}/a$ is the conformal Hubble rate. The symbol $\bar{\rho}$ denotes the mass density of baryons and dark matter, and $\alpha$ and $\gamma$ are PPN parameters that are interpreted as determining the value of Newton's constant and the curvature of space per unit mass, respectively \cite{Will_1993}. The additional parameters $\alpha_c$ and $\gamma_c$ are neglected in the Solar System, but are needed here in order to account for dark energy \cite{Sanghai_2017} (including any component corresponding to the cosmological constant, $\Lambda$).

The Fourier transforms of the perturbation equations governing $\Phi$ and $\Psi$ are \cite{Sanghai_2017,Sanghai_2019} 
\begin{eqnarray}
-\mathcal{H}^2 \Phi- \mathcal{H} \Psi_{,\tau} - \frac{1}{3} k^2 \Psi &=& -\frac{4\pi G \bar{\rho} a^2}{3} \, \mu \, \delta  \label{eq_pert1}\\
2\mathcal{H}_{,\tau}\Phi+\mathcal{H}\Phi_{,\tau}+\Psi_{,\tau\tau} +\mathcal{H}\Psi_{,\tau} - \frac{1}{3} k^2\Phi &=& -\frac{4\pi G  \bar{\rho} a^2 }{3} \, \xi \, \delta \, , \label{eq_pert2}
\end{eqnarray}
and \cite{Anton_2021}
\begin{equation}
\Psi_{,\tau} +\mathcal{H}\Phi=4\pi G a^2  \, \mu \, \bar{\rho} \, v +\mathcal{G} \, \mathcal{H} \Psi\label{eq_pert3} \, \text{,}
\end{equation}
where $k$ is the comoving wavenumber, $\delta \equiv \delta \rho/\bar{\rho}$ is the density contrast, and $v$ is the velocity potential (defined implicitly by $v_i\equiv v_{,i}$, where $v_i$ is the 3-velocity of matter\footnote{The reader may note that the factor $\bar{\rho} v$ in Eq. (\ref{eq_pert3}) can be naturally extended to $\rho v_i$ in the non-linear regime, so that the appropriate Newtonian limit is fully recovered \cite{Sanghai_2017}.}). To complete the set of equations it is useful to also define a `slip' $\eta \equiv \Phi / \Psi$ \cite{caldwell2007constraints, amendola2008measuring, Thomas_2023}.

The functions $\mu$, $\xi$, $\mathcal{G}$ and $\eta$ are coupling functions that incorporate the effects of modifying gravity, and which have a scale dependence of the following form\footnote{We note the similarity of this approach to the use of Pad\'{e} approximants to model the scale dependence of gravitational couplings, as in Refs. \cite{amin2008subhorizon, bakerbull}.} \cite{Thomas_2023}:
\begin{equation}
\label{eqn_timinterp0}
f(k)=\frac{1}{2} \left( f_{\rm S}+f_{\rm L} \right) + \frac{1}{2} \left( f_{\rm S}-f_{\rm L} \right) \tanh\left( \ln \frac{k}{\mathcal{H}}  \right)\, ,
\end{equation}
where $f$ in this equation is intended to represent any one of the set $\{ \mu, \xi, \mathcal{G}, \eta \}$, and where $f_{\rm S}$ and $f_{\rm L}$ are the small and large-scale parts of this quantity. For $\{ \mu, \xi, \mathcal{G} \}$ we have
\begin{eqnarray}
\mu_{\rm S}&= \gamma \, , \qquad \qquad \qquad \mu_{\rm L}&= \gamma -\dfrac{1}{3} \hat{\gamma}+\dfrac{1}{12\pi G \bar{\rho}} \, \hat{\gamma}_c\, , \label{eq_ppnc_mu} \\
\xi_{\rm S}&= \alpha \, , \qquad \qquad \qquad \xi_{\rm L}&= \alpha - \dfrac{1}{3}\hat{\alpha}+\dfrac{1}{12\pi G \bar{\rho}} \, \hat{\alpha}_c\, , \label{eq_ppnc_xi} \\
{\mathcal{G}}_{\rm S}&= \dfrac{\alpha-\gamma}{\gamma}+ \dfrac{\hat{\gamma}}{\gamma}  \, , \qquad {\mathcal{G}}_{\rm L}&= 0 \, , \label{eq_ppnc_curlyG}
\end{eqnarray}
where hats denote derivatives with respect to the number of e-foldings, i.e. $\hat{X} \equiv \mathrm{d}X/\mathrm{d}(\ln \, a)$. The slip on small scales is given by $\alpha/\gamma$, but a corresponding expression for the large-scale limit of this quantity has so far proven to be elusive. These equations fully specify both the background and the scalar perturbations in terms of the extended set of PPN parameters $\{ \alpha , \gamma , \alpha_c, \gamma_c \}$. General-relativistic $\Lambda$CDM cosmology is recovered when $\alpha=\gamma=1$ and $\alpha_c=-2 \gamma_c= \Lambda$, such that $\mu=\xi=\eta= 1$ and $\mathcal{G}=0$.

We note that all four of the set $\{ \alpha , \gamma , \alpha_c, \gamma_c \}$ are being treated here as being functions of time, but not space. The time-dependence that has been allowed for these functions is essential for their use in cosmology, and failure to include it would constitute a severe restriction on the theories of gravity that it could be used to describe. Spatial independence of these parameters, on the other hand, is an important feature of the PPN approach, and has been retained. We note that we still refer to $\{ \alpha , \gamma , \alpha_c, \gamma_c \}$ as ``parameters'', despite their time-dependence, for linguistic ease as well as for continuity with the existing literature.

We note also that the four parameters $\{ \alpha , \gamma , \alpha_c, \gamma_c \}$ are not all independent, but must instead obey the integrability condition
\begin{equation} \label{eq_integrability}
4 \pi G\,\bar{\rho} \left( {\alpha-\gamma+ \hat{\gamma}} \right)= {\alpha_c + 2 \gamma_c + \hat{\gamma}_c} \, .
\end{equation}
This condition reduces the number of functional degrees of freedom that are required down to three. However, the absence of an expression of the large-scale slip in terms of the other parameters requires the specification of one additional function of time (but not scale). This cancels the reduction in functional degrees of freedom provided by the integrability condition, at least until a general expression for $k\rightarrow 0$ limit of $\eta$ is found. Finally, let us note that the equations above are valid for (i) `fully conservative' theories of gravity \cite{Will_1993}, (ii) spatially flat cosmologies, and (iii) for dust-domination only. The first two of these can be straightforwardly relaxed, and are made here only to keep the presentation concise. The last is a condition of the post-Newtonian  expansion used in building this framework (see, however, Ref. \cite{Sanghai_2016}).

The equations above are remarkable in that they describe the evolution of both the cosmological background and the perturbations in terms of the (extended) PPN parameters only. They therefore describe the cosmological behaviour of any theory of gravity that fits into the PPN framework\footnote{Note, however, that theories of gravity that involve screening mechanisms are as yet not included in this approach, as they do not directly fit into the standard PPN framework. However see e.g. \cite{Avilez_Lopez_2015,ppnv2, ppncscreen} for a possible route to their inclusion.}. Even more remarkably, the density contrast is not required to be small in this approach, meaning that it can used to describe non-linear structure formation without any additional parameterisation. For the observables we use in this work, we will not be using this property directly, as linear theory will be sufficient. We leave the examination of the PPNC phenomenology in the non-linear regime to future work.

\subsection{Code implementation} \label{subsec:code_implementation}

We implement the framework described above by modifying the Cosmic Linear Anisotropy Solving System (CLASS) \cite{lesgourgues2011cosmic}, which is a Boltzmann code that evolves perturbations from early times up to the present day. In particular, we implement the modified Friedmann equation, the momentum constraint equation (used to evolve $\Psi$) and the slip equation (used to calculate $\Phi$).\footnote{We note that the use of the slip to reduce the number of required constraints when evolving the PPNC equations was shown in \cite{Thomas_2023}.} See Appendix \ref{app_code} for full details of the implemented equations. In deploying our approach we necessarily need to make some choices, which we detail here. They are: (i) a choice for how to include radiation, and (ii) a choice of how to specify the functional form of the time dependence of the parameters $\{\alpha(\tau), \gamma(\tau), \alpha_c(\tau), \gamma_c(\tau) \}$, and the large-scale limit of the slip $\eta_L(\tau)$, which are not uniquely specified by the PPNC approach itself. 

The PPNC framework outlined above is constructed by considering the gravitational fields of pressureless dust and dark energy, and does not explicitly include radiation or neutrinos in its present form. This is due to the PPN formalism having been constructed to deal with the gravitational fields of astrophysical bodies in the late Universe, which have mass density as the leading-order contribution to the source of their gravitational fields. This is not true of the early Universe, where relativistic species such as radiation and neutrinos have an isotropic pressure that also contributes at leading order. In principle, this situation could be remedied by extending the PPN formalism to include such contributions, but we expect this to be a non-trivial task that may involve generalizing the post-Newtonian expansion itself, as well as potentially involving the introduction of additional coupling parameters (see Ref. \cite{Sanghai_2016} for preliminary steps). 

For our implementation in the present study we have made what we consider to be the most conservative choice possible; not modifying the gravitational coupling to other species at all, such that (for example) the Friedmann equation in the code appears as
\begin{equation}\label{eqn_Friedmann_code}
    H=\sqrt{\frac{8\pi G}{3}\left(\bar{\rho}_\text{tot}-\bar{\rho}_c-\bar{\rho}_b\right)+\frac {8\pi G}{3}\left(\bar{\rho}_c+\bar{\rho}_b\right)\gamma-\frac{2\gamma_c}{3}}
\end{equation}
where we have taken the curvature of space to vanish, and where $\bar{\rho}_\text{tot}$ is the total energy density, and $\bar{\rho}_c$ and $\bar{\rho}_b$ are respectively the energy densities of cold dark matter and baryonic matter, with the source terms of other equations being prescribed similarly. This choice is not unique, but provides us with a definite means to calculate the effect of the change in gravitational coupling strengths of pressureless matter without having to make significant changes to the radiation-dominated stage of the Universe's history.

\begin{figure}
    \hspace{0.5cm}
    \includegraphics[width=0.8\linewidth]{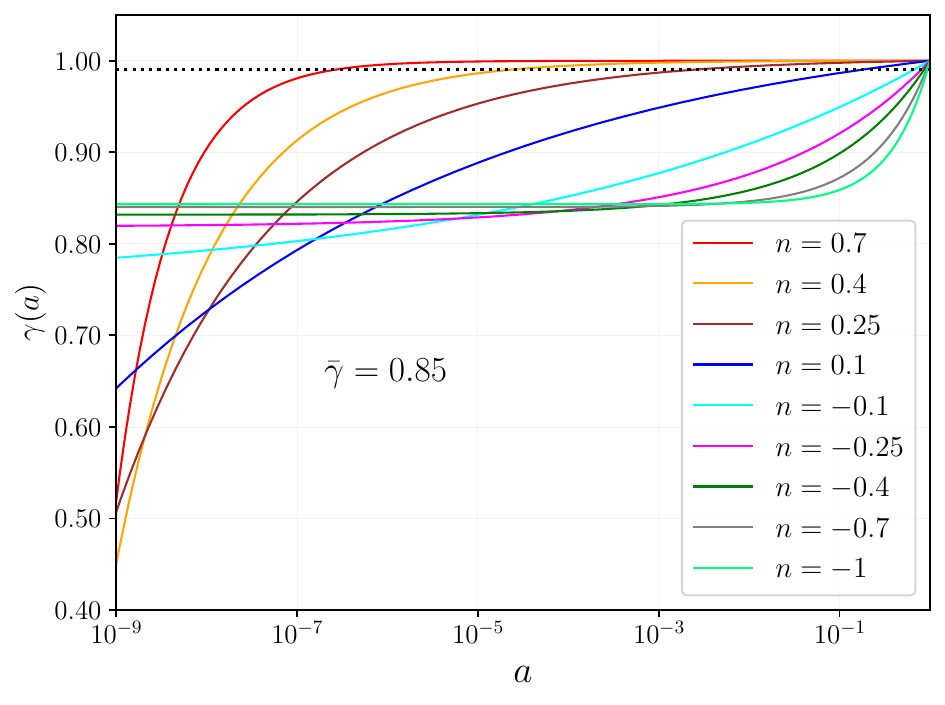}
    \caption{Evolution of the PPN parameter $\gamma(a)$ for different values of the power-law index $n$, shown as a function of the scale factor $a$. In each case shown, $\bar{\gamma}$ is equal to $0.85\,$. The dotted line is at $\gamma = 0.99\,$, showing that for larger values of the power law ($n=0.4$, or $0.7$, for the value of $\bar{\gamma}$ used here), the deviations from GR are small by the time that matter contributes significantly to the evolution of the background cosmology.}
    \label{fig_gamma_of_a_power_laws}
\end{figure}

We choose to specify the time-dependence of $\alpha$ and $\gamma$ by making them functions of the scale factor $a$, in the following way:
\begin{eqnarray} \label{eq_ppnc_power_law_n}
\alpha(a)=A\left(\frac{a_1}{a}\right)^n+B \qquad {\rm and} \qquad    \gamma(a)=C\left(\frac{a_1}{a}\right)^n+D\text{,}
\end{eqnarray}
with the constants $\{A, B, C, D\}$ calculated from the values of $\alpha$ and $\gamma$ at the initialization time $a=a_1$ and at the present time $a=1$ (which is conventionally equal to unity for $\alpha$, and strongly constrained from astrophysical observations for $\gamma$ \cite{Will_2014,Bertotti_2003}). The value of $n$ can then be used to control the rate of change of the parameters over cosmic history, and in what follows we will consider different values for this quantity. The power-law forms for $\alpha$ and $\gamma$ are chosen for their simplicity, and the smoothness of the functions that result (the time derivatives of these functions are a crucial part of the expressions for the coupling functions). Treating them as piece-wise functions in bins would not result in the required smoothness, and tying them to dark energy (as in e.g. Ref.  \cite{refId0}) would not appear to be a sensible choice in the present context, as we are interested in precision tests of gravity rather than testing candidates for dark energy. Our power-law approach is supported by the behaviour found in Ref. \cite{Thomas_2023} for some example theories of gravity, where in those cases the same power-law index\footnote{A value of $n$ close to 0.1 was a good fit for the theories in Ref. \cite{Thomas_2023} that were furthest from $\Lambda$CDM, which is partly why we choose $n=0.1$ in some of our investigations below.} for both $\alpha$ and $\gamma$ was found to be a good description, which supports the choice here to use the same power-law index for both $\alpha$ and $\gamma$. This choice also aids the simplicity of this initial study, but could be easily relaxed if desired. We anticipate that Gaussian processes or genetic algorithms might be better and more model-independent ways to reconstruct the PPNC functions from observations, but we leave the use of these more sophisticated approaches for future work. The behaviour of $\gamma(a)$ for different values of the power-law index $n$ is displayed in Fig. \ref{fig_gamma_of_a_power_laws}.

The value of $\gamma_c$ at $z=0$ is determined by its fractional contribution to $H$ at this time (see Appendix \ref{app_code}). For this study we choose to set $\gamma_c$ to a constant, meaning that is fully specified by its value at $z=0$, thereby allowing us to focus on the phenomenology of $\alpha$ and $\gamma$. In addition, this choice makes the $\gamma_c$ contribution to the background as similar as possible to $\Lambda$, which is well understood. We use the integrability condition to specify $\alpha_c$ at each time, so specifying the time dependence of this parameter is not required (in fact for the choice of constant $\gamma_c$, we find $\alpha_c$ will also be close to constant). The choice to specify $\gamma_c$ directly, and $\alpha_c$ from the integrability condition, has the effect of making the background expansion dependent only on $\gamma$ and $\gamma_c$. We could alternatively specify $\alpha_c$ directly, but doing so requires an additional value to be provided, so here we instead choose to use the minimal approach of specifying $\gamma_c$ directly.

In general, our code implementation allows a non-GR value of the large-scale slip, $\eta_{\rm L}$, to be specified as an additional (time-varying) degree of freedom. However, as discussed earlier, we choose to set it equal to its general-relativistic value of one for this initial study. This can be easily relaxed if desired, or re-visited if/when a theoretical prediction of this quantity is found within the PPNC framework. The small-scale slip is implemented as $\eta_{\rm S} = \alpha/\gamma$, as required by the PPNC equations. For all of the PPNC modifications, we choose to take them as not being present when the initial conditions are set, but rather to switch them on deep within the radiation-dominated era, at $a_1=10^{-10}$, where their contributions are negligible.

With the choices made here, the PPNC degrees of freedom to be constrained later are $\gamma(a_1)$, $\alpha(a_1)$ and their power-law index $n$. In order to reduce the degeneracy between the first two parameters and their power law, and work with more physically relevant variables, we present our results using the averages of $\alpha$ and $\gamma$ over cosmic time, defined as
\begin{equation} \label{agbar}
\bar{\alpha}\equiv \frac{\int^{0}_{\ln a_1}{\alpha(a)\,\mathrm{d} \ln a}}{\int^{0}_{\ln a_1}{\mathrm{d} \ln a}}
\qquad {\rm and} \qquad
\bar{\gamma}\equiv \frac{\int^{0}_{\ln a_1}{\gamma(a)\,\mathrm{d} \ln a}}{\int^{0}_{\ln a_1}{\mathrm{d} \ln a}}\text{.}
\end{equation}
Two additional derived parameters calculated by the modified Boltzmann code are presented in the MCMC results, which are the derivatives of $\gamma$ and $\alpha$ with respect to the scale factor at the present time ($a=1$), denoted $\gamma'(a)$ and $\alpha'(a)$.

The full set of perturbation, background, PPNC functions, and constraint equations, as implemented in the code, and in the notation and conventions used within CLASS, are given in Appendix \ref{app_code}.

\subsection{Relationship to varying $G$}
Constraints on the variation of Newton's constant, $G$, are typically not equivalent to the constraints we seek to impose on $\alpha$ and $\gamma$, as they tend to modify only a specific equation in which $G$ appears, and usually only in a particular context. Relativistic theories of modified gravity can, however, effectively modify $G$ differently in different contexts, so that constraints on $G$ in one context do not necessarily apply in other contexts. As such, constraints on the variation of $G$, and equivalently cosmological tests using $\mu$ as a modification of the Poisson equation in cosmological perturbation theory, are not generally equivalent\footnote{See e.g. Ref. \cite{ppnvscosmo} for related comments on this issue.} to constraints on $\alpha$ and $\gamma$. A key exception to this are the Solar System tests of $\dot{G}$, which occur in the same weak-field expansion about Minkowski space as the PPN framework. The constraints from these tests are therefore equivalent to a constraint on $\dot{\alpha}$ at $z\approx0$. A good summary of these tests, and the associated references, is given in Ref. \cite{Will_2014}, with the Mars ephemeris constraints being the strongest, giving $\dot{\alpha}
=0.1\pm1.6 \times 10^{-13} \, {\rm yr}^{-1}$ (assuming mass loss from the sun is negligible). This is followed by lunar laser ranging (about five times worse) and helioseismology (around an order of magnitude worse). Conversely, the primordial nucleosynthesis constraints cited in the same reference\footnote{See e.g. Ref. \cite{bbn} for a more recent treatment.} do not correspond to the time-variation of the PPN parameters $\alpha$ or $\gamma$ at all, as they derive  constraints from modifying the background Friedmann equation during the radiation-dominated era, rather than the coupling parameters for matter fields.

\subsection{MCMC methodology}\label{sec:mcmc}
Our constraints are obtained using a Markov Chain Monte Carlo (MCMC) approach, using the publicly available MontePython code \cite{audren2013conservative,audren2013monte}. MontePython calls the modified PPNC CLASS code through a Python wrapper; for all runs we use the Metropolis Hastings option within MontePython and a jumping factor of 2.1.

Seven standard cosmological parameters are varied in our MCMC analyses: the dark matter dimensionless density $\omega_c$, the baryon dimensionless density $\omega_b$, the Hubble parameter $H_0$ (in units of km/s/Mpc), the optical depth $\tau$, the quantity $\ln(10^{10}A_s)$ where $A_s$ is the amplitude of scalar perturbations, the spectral index of scalar perturbations $n_s$, and the primordial helium fraction $Y_p$\,. We note that although the last of these is not independent of the first six parameters in a $\Lambda$CDM context, that is no longer true in the present case as the background expansion can be affected in the radiation era, which necessitates that this parameter is varied independently in the MCMC analysis. The standard parameters that were not varied in our analysis were set as follows: the spatial curvature was set to zero, as was the cosmological constant $\Omega_\Lambda$ (with dark energy having been absorbed into $\alpha_c$ and $\gamma_c$, as described above). We used two massless neutrinos, and one with mass $0.06 \,{\rm eV}$, keeping the effective number of neutrinos to $N_\text{eff} = 3.046$. We also note that we include adiabatic perturbations only.

In all of our runs we vary the time-averaged PPN parameters $\bar{\alpha}$ and $\bar{\gamma}$. We perform some runs with the power-law index $n$ from Eq. (\ref{eq_ppnc_power_law_n}) fixed to different values, and some runs with $n$ as a free parameter in order to marginalise over it. As we are focused on a simple initial exploration of the constraints that can be obtained, we do not vary the other PPNC parameters in the code for this work. In particular, in all cases we set $\alpha(a=1)=1.0$ (as per the standard convention), $\gamma(a=1)=1.0$ (in line with Solar System constraints\footnote{A more comprehensive approach would be to include the constraint on $\gamma(a=1)$ as a Gaussian prior, however the variance of this prior from Solar System tests would be very close to the GR value, and much smaller than the constraining power of cosmology, so we adopt this simpler approach for this initial analysis.}), and we set the large-scale slip equal to one (its GR value). We keep $\gamma_c$ constant in time, as explained in the previous section, meaning that it is fully specified by the Friedmann equation at $z=0$ (see Appendix \ref{app_code}). We use the default start time for the PPNC effects to be when the scale factor takes the value $a=10^{-10}$ (the precise value being unimportant here, as long as it occurs deep within the radiation era). We leave an exploration of the full freedom from varying all of these parameters simultaneously to future work.

The main dataset that we use is the Planck 2018 data release \cite{Planck_2020}, comprising the low-$\ell$ likelihood, the full TT, EE and TE high-$\ell$ likelihood with the complete ``not-lite'' set of nuisance parameters, and the Planck lensing potential likelihood (see e.g. \cite{planck_likelihood} and the Planck legacy archive\footnote{\href{https://wiki.cosmos.esa.int/planck-legacy-archive/index.php/Main_Page}{https://wiki.cosmos.esa.int/planck-legacy-archive/index.php/Main\_Page}} for full details of these likelihoods). We include Gaussian priors on the nuisance parameters (also varied as MCMC parameters) as recommended by the Planck collaboration \cite{planck_likelihood} and implemented in the MontePython code. We include flat priors on the standard cosmological parameters, and flat non-negative priors on $\gamma(a_1)$ and $\alpha(a_1)$ (i.e. $0\le\alpha(a_1),\gamma(a_1)\le50$). We use purely linear modelling of the observables, as non-linear extensions have not yet been validated within the PPNC approach and non-linearities only have a small effect on CMB lensing at the level of the Planck data.\footnote{We verified that non-linearities are subdominant for our results by running with halofit switched on; the resulting changes to the constraints were negligible. However, constraints from future experiments such as the Simons Observatory \cite{Ade_2019}, are likely to require non-linear modelling for the lensing potential.}

For the runs with varying power-law index $n$, we apply a flat prior with a lower bound $-15<n$, but we note that the results depend on the upper bound of this prior, $n_\text{max}$. The reason for this is shown in Fig. \ref{fig_gamma_of_a_power_laws}: increasing the value of $n$ moves the consequences of modifying gravity to earlier and earlier times, deeper and deeper into the radiation era. At these times the matter fraction is increasingly small and thus the parameters have very little effect regardless of the starting values of $\alpha$ and $\gamma$, because the PPN parameters have been assumed to not couple to radiation; we refer to this region of parameter space as the {\it physically uninteresting region}. We run three sets of chains, with the upper bounds on the prior for $n$ set to $0.25$, $0.4$ and $1.0$. The last of these is very conservative, as it allows all of the effects of modifying gravity to be confined to very early times. In this case the posterior is focused in the range of values of $n$ where the altered gravitational parameters have essentially zero effect, allowing for unrealistically large values of $\gamma$ and $\alpha$.

The value $n_\text{max}=0.25$ was chosen as a low value of the prior that includes the vast majority of the physically interesting parameter space. This value was reached by examining the extent to which the deviation of $\gamma$ and $\alpha$ from their GR value has dropped by matter radiation equality for each power-law, i.e. $({\gamma(a_{eq})-1})/({\gamma(a_\text{start})-1})$. The power-law at which this ratio is approximately $0.02$ is $n=0.25$, which means that for this power-law, the modification to the Friedmann equation at the redshift of matter-radiation equality\footnote{Taken as the redshift in a $\Lambda$CDM cosmology with identical values of the non-PPNC parameters.} is around $1\%$ of the deviation of the PPN parameters from GR at $a_\text{start}$. For $n=0.2$ this ratio is approximately $0.04$, and for $n=0.3$ it is approximately $0.01$. We believe that the value of this ratio for $n=0.25$ ensures a sensible minimum contribution from the modified gravitational parameters. In other words, an upper bound of $0.25$ on this prior is equivalent to stating that the deviation of the PPN values from unity (their GR value) cannot drop to less than $1\%$ of the starting difference before matter-radiation equality. We comment on the reasonableness of this value later on, given the results we obtain. The third value, $n=0.4$, was chosen from the fixed power-law runs as the one for which the end of the transition region has been reached. This run acts as a sanity check on the choice of upper prior $n=0.25$. For this power-law, we find $({\gamma(a_{eq})-1})/({\gamma(a_\text{start})-1}) \simeq 2.5\times 10^{-3}$, which is around an order of magnitude smaller than for the $n=0.25$ case.

For the fixed power-law runs we obtain a suitable covariance matrix from some exploratory runs, and then we use ten chains for each run, continuing until the Gelman-Rubin `$1-R$' convergence test \cite{gelman1992inference} is smaller than $0.01$ for all parameters. We run chains for $n \in \{-1, -0.7, -0.4, -0.1, 0.1, 0.25, 0.4, 0.55, 0.7, 1\}$.

The posteriors for the varying power-law cases are more complicated, so we follow a slightly more involved procedure to ensure convergence. The ``T-shaped'' degeneracy that occurs in this case (see later), and the broader allowed range of the PPN parameters for large values of the power-laws due to the physically uninteresting region, mean that the covariance matrix around the central value of the posterior is ill-suited as a proposal distribution for smaller values of the power-law $n$ (it is too broad). As such, we start exploratory chains around the best-fit point for the fixed $n=0.1$ power law case (this is the highest likelihood point we found with significant non-GR behaviour), and use these exploratory chains to get a (narrower) covariance matrix that gives us a sensible overall acceptance rate (i.e. $0.22$, which is similar to the fixed power-law case) and that can be seen to sample sufficiently when chains are exploring lower values of the power-law. We then run around $120$ chains (starting from the same $n=0.1$ best-fit point) of around the same length of the $10$ required for the fixed power-law cases to converge, and which results in a typical number of steps of more than $600,000$ per chain (ranging from $\sim 440,000-770,000$). The chains are individually inspected for converged behaviour, and the large number of chains ensures that the parameter space is well explored and not dependent on the behaviour of only a small number of individual chains.

\section{Results}

We present results for both fixed and varying power-law chains, before examining the physical causes of the trends that are found in the constraints. We note that this is a nested model containing $\Lambda$CDM, and that in all cases we find no significant preference for non-GR values of the PPN parameters, such that we see no need for a formal model comparison.

\subsection{Fixed power-laws}

We first examine the constraints on the PPN parameters when the power-law index for $\alpha$ and $\gamma$ is fixed; we consider this for $n \in \{-1.0, -0.7, -0.4, -0.1, 0.1, 0.25, 0.4, 0.55, 0.7\}$.

\subsubsection{Constraints on $\bar{\alpha}$ and $\bar{\gamma}$}

\begin{figure}
\hspace{-1.5cm}
\includegraphics[width=1.2\linewidth]{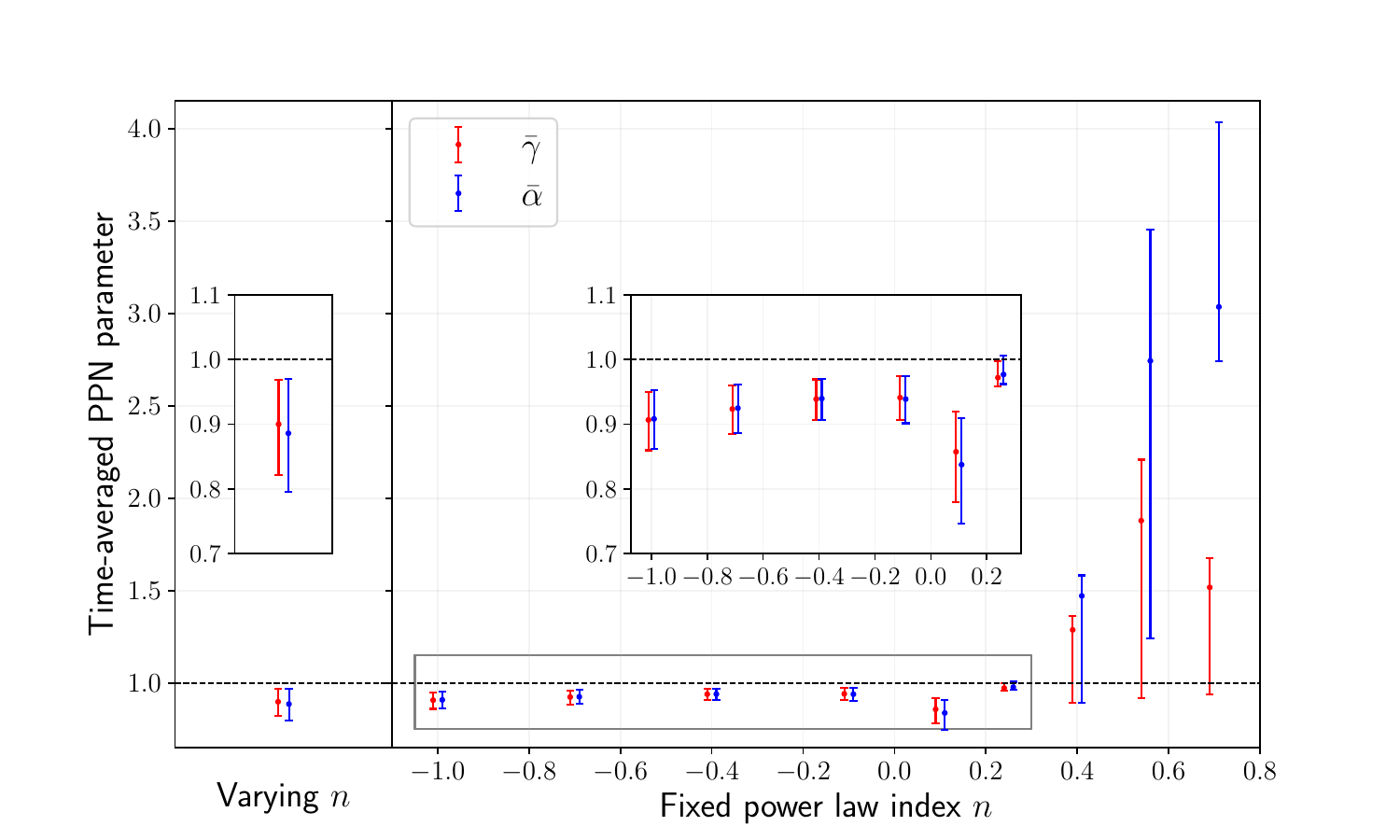}
\caption{Constraints on $\bar{\gamma}$ and $\bar{\alpha}$ for fixed power-law with different values of $n$, and for the case where $n$ is an additional free parameter with a flat prior $n \in [-15, 0.25]$. The constraints on $\bar{\gamma}$ and $\bar{\alpha}$ are very similar, and the results are similar for most values of $n$ as well as in the varying power-law case. A change in behaviour for power-laws with $n>0.25$ is clearly visible. The error bars show the $68\%$ confidence interval.}
\label{fig_gammaalphaerror}
\end{figure}

Our main results are displayed in Table \ref{table_alphagamma}, where we show the 68\% confidence intervals on $\bar{\alpha}$ and $\bar{\gamma}$, for fixed power-laws with different values of $n$. The same information is shown graphically in Fig. \ref{fig_gammaalphaerror}.

The PPN parameters are constrained very similarly, due to the strong degeneracy that occurs between them, as shown in Fig. \ref{fig_h0omcerror2d}.  For most values of the power-law in the interval $-1<n<0.25$, the two parameters are constrained to within $14\%$ of their GR value of unity. The width of the posteriors is about $9\%$ of their GR value, with the best constraints being given for the $n=0.25$ case, where the width of the posteriors are about $4\%$ of their GR value. The behaviour is a little different at $n=0.1$, where the constraints are 1.5-2 times worse and the PPN parameters are allowed to differ from their GR values by up to 25\%; this is also where the mean is furthest from its GR value. This is due to the various degeneracies involved with this power law, both between the PPN parameters and between the PPN and $\Lambda$CDM parameters, which allow for a wider range of PPN parameters to be well compensated. Interestingly, this is the power law that is most similar to the behaviour found in an earlier study that considered specific theories \cite{Thomas_2023}. We discuss all of this phenomenology in more detail in Sections \ref{sec_phenomenology} and \ref{sec_interpretation}. Table \ref{table_alphagamma} also shows that there is a mild preference for weak gravity: slightly better values of the likelihood can be found for values of $\bar{\alpha}$ and $\bar{\gamma}$ less than unity. This improvement comes from small improvements to each of the low-$\ell$, high-$\ell$ and lensing likelihoods.

As the power-law gets larger, the trends identified above change: the constraints get worse and increasingly one-sided, with a central value that gets larger and larger. This is due to the previously identified issue that for larger power-laws, the modified gravity effects are increasingly confined to deeper and deeper within the radiation era, and thus the effects on the cosmology are much smaller (for the same amplitude of PPN parameter values). The constraints on $\bar{\alpha}$ and $\bar{\gamma}$ are less quantitatively similar for these power-laws. In this case the constraints become one-sided due to the non-negativity of the prior on the starting values of $\alpha$ and $\gamma$ causing there to be a much larger volume of parameter space for PPN parameters greater than one. The large increase in allowed range of the PPN parameters is further shown in the 2D posteriors displayed in Fig. \ref{fig_h0_alphagamma_2d_2}.

We present the $95\%$ confidence intervals in Appendix \ref{app_constraints}. In all cases the parameters are constrained to lie within $30\%$ of their GR values at the $95\%$ confidence level, and in all cases except one the results are consistent with the general-relativistic $\Lambda$CDM cosmology. The exceptional case is $\bar{\alpha}$ for $n=0.1$: in this case there is a $3\%$ deviation from GR at the $95\%$ confidence level, however even this case is consistent with GR at the $99.7\%$ confidence level. Comparison of the $68\%$ and $95\%$ confidence-level constraints shows the clearly non-Gaussian behaviour of the constraints for larger values of the power-law index $n$.

\begin{table}
\hspace{-1.4cm}
\begin{mytabular}[1.2]{|c|c|c|c|c|c|c|c|c|} 
\hline 
Power-law & $-1$ & $-0.7$ & $-0.4$ &  $-0.1$ & $0.1$ & $0.25$ & $0.4$ & varying \\
\hline 
$\bar{\gamma}$ & $0.91^{+0.04}_{-0.05}$ & $0.92^{+0.04}_{-0.04}$ & $0.94^{+0.03}_{-0.03}$ & $0.94^{+0.03}_{-0.03}$ & $0.86^{+0.06}_{-0.08}$ & $0.97^{+0.03}_{-0.01}$ & $1.29^{+0.08}_{-0.40}$ & $0.90^{+0.07}_{-0.08}$ \\
\hline 
$\bar{\alpha}$ & $0.91^{+0.04}_{-0.05}$ & $0.92^{+0.04}_{-0.04}$ & $0.94^{+0.03}_{-0.03}$ & $0.94^{+0.04}_{-0.04}$ & $0.84^{+0.07}_{-0.09}$ & $0.98^{+0.03}_{-0.01}$ & $1.47^{+0.11}_{-0.58}$ & $0.89^{+0.08}_{-0.09}$ \\
\hline
\end{mytabular} 
\caption{Constraints on $\bar{\gamma}$ and $\bar{\alpha}$ for fixed $n$, and for varying $n$ (with $n_{\rm max} = 0.25$ as the upper bound on the prior), with their $68\%$ credible regions.}
\label{table_alphagamma}
\end{table}

\subsubsection{$\Lambda$CDM parameters}

\begin{figure}
\hspace{-1cm}
\includegraphics[width=1.1\linewidth]{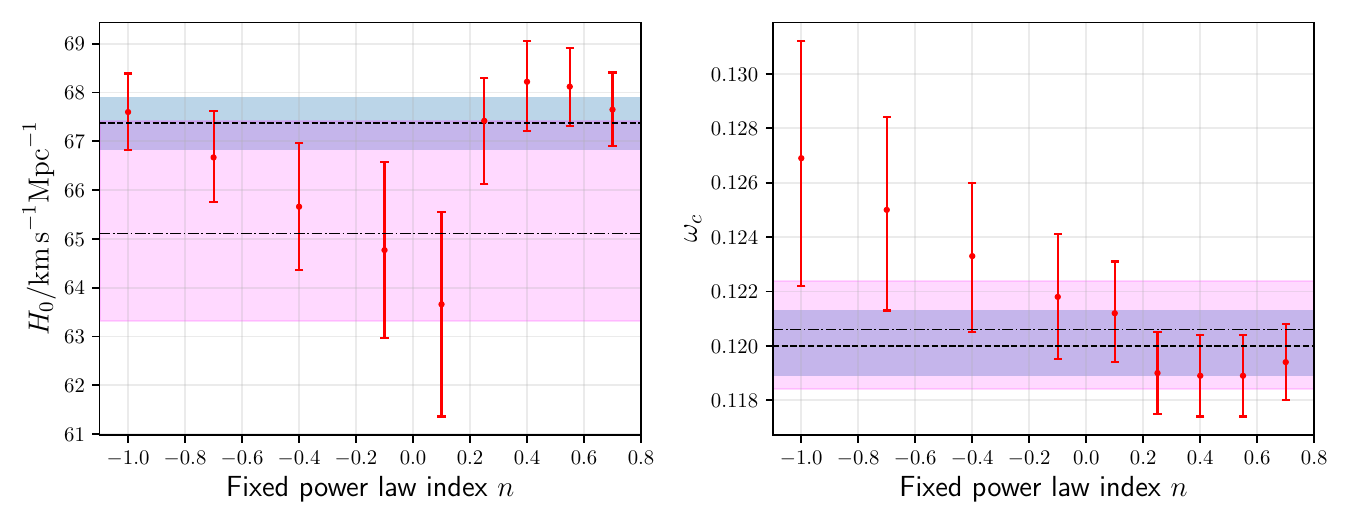}
\caption{Constraints on $H_0$ and $\omega_c$\,, for different values of the fixed power-law index, $n$. The blue shaded regions are the constraints on these parameters for the 7 parameter $\Lambda$CDM model (including $Y_p$), with the mean value given by the dashed black line. The magenta shaded regions are the constraints for the varying power-law case discussed in Section \ref{subsec:varying_powerlaw}, with the mean value given by the dot-dashed black line.}
\label{fig_h0omcerror}
\end{figure}

We find that most $\Lambda$CDM parameters have at least minor degeneracies with the PPN parameters, and thus the ability of the data to constrain them degrades. Notably, there is a 20-30\% worsening of the constraints on $\omega_b$ and $Y_p$, which does not significantly change with power-law. The more interesting changes are for $H_0$ and $\omega_c$; the resultant constraints on these parameters are shown in Fig. \ref{fig_h0omcerror}.

\begin{figure}
\centering
\hspace{-1.5cm}
\begin{subfigure}{0.55\textwidth}
    \includegraphics[width=\linewidth]{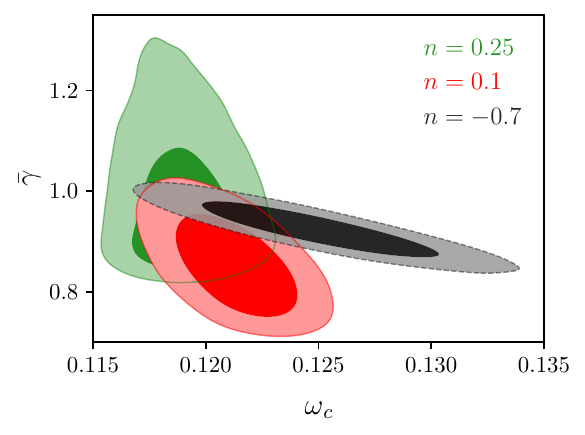}
    \label{fig_omcerror}
\end{subfigure}
\hspace{-0.8cm}
\begin{subfigure}{0.55\textwidth}
    \includegraphics[width=\linewidth]{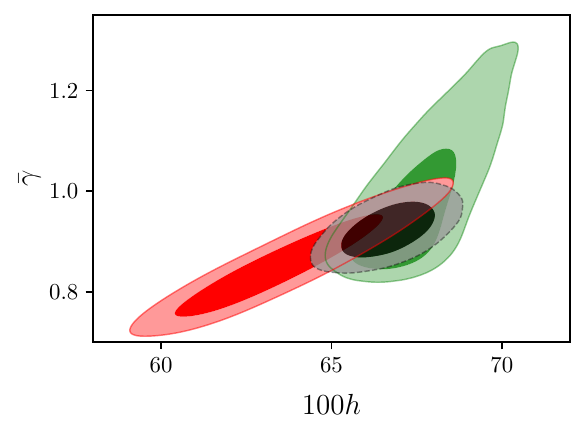}
    \label{fig_h0error}
\end{subfigure}
\vskip -1.5ex
\begin{subfigure}{0.55\textwidth}
    \includegraphics[width=\linewidth]{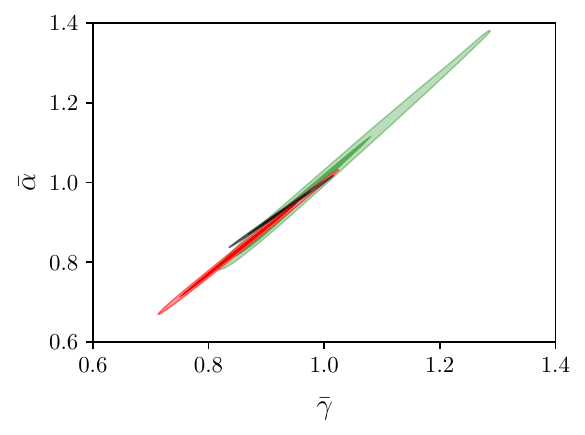}
    \label{fig_gammaerror}
\end{subfigure}
\caption{Depiction of how the 2D posteriors change for different fixed power-law index $n\,$. The top left plot shows the $\left(\omega_c,\bar{\gamma}\right)$ degeneracy, the top right shows $\left(H_0,\bar{\gamma}\right)$ (with $H_0$ expressed in terms of the dimensionless $h$: $H_0 = 100 h\, {\rm km}\,{\rm s}^{-1}\,{\rm Mpc}^{-1}$) and the bottom shows $\left(\bar{\gamma}, \bar{\alpha}\right)\,$. The values of power-law index have been to demonstrate how the degeneracies change with $n$. The $\bar{\gamma}=\bar{\alpha}$ degeneracy is very strong in all cases. The $H_0$ degeneracy is extended for $n=0.1$ compared to most of the power-law range. The $\omega_c$ degeneracy gets stronger for smaller power laws, as a larger change is required to compensate the effects caused by the PPN parameters. In all cases the allowed range of the PPN parameters extends for the $n=0.25$ case, as their effects are pushed to earlier and earlier times.} 
\label{fig_h0omcerror2d}
\end{figure}

The constraint on $\omega_c$ is similar to its value in $\Lambda$CDM for larger values of the power-law index ($n=0.7$), and gets monotonically worse for smaller power-laws, becoming 3-4 times worse for a power-law index of $n=-1.0$. This is due to a changing degeneracy between $\omega_c$ and the residual effect of the PPN parameters, after the cancellation that causes the $\bar{\gamma}$ - $\bar{\alpha}$ degeneracy. This degeneracy is shown by the 2D posteriors in the top left plot of Fig. \ref{fig_h0omcerror2d}, and is discussed in more detail in Sections \ref{sec_phenomenology} and \ref{sec_interpretation}. The constraint on $H_0$ is similar to $\Lambda$CDM for the more extreme power-law values (i.e. $-1.0$ and $0.7$), but is up to three times worse for values in between. This is because the degeneracy between $H_0$ and the residual effect of the PPN parameters, after the cancellation that causes the $\bar{\gamma}$ - $\bar{\alpha}$ degeneracy, is strongest within this region. This degeneracy is also the cause of the central value being pulled away from its $\Lambda$CDM value in this range (the $\Lambda$CDM value is not within the 68\% credible region for a power-law of $n=0.1$). We note that this is not a significant effect, however, and it is in the wrong direction to help resolve the $H_0$ tension \cite{di2021realm}. This degeneracy is shown by the 2D posteriors in the top right plot of Fig. \ref{fig_h0omcerror2d}, and is shown in more detail in Fig. \ref{fig_H0_omc_gamma_corner_both}. It is also discussed in more detail below.

For both $H_0$ and $\omega_c$, the error bars for large values of $n$ tend towards their $\Lambda$CDM values as the effects of the PPN parameters are pushed towards earlier times, and the degeneracies are increasingly weakened.

\subsubsection{Constraints on $\alpha'$ and $\gamma'$ at $z=0$}

\begin{figure}
\hspace{-1.5cm}
\includegraphics[width=1.2\linewidth]{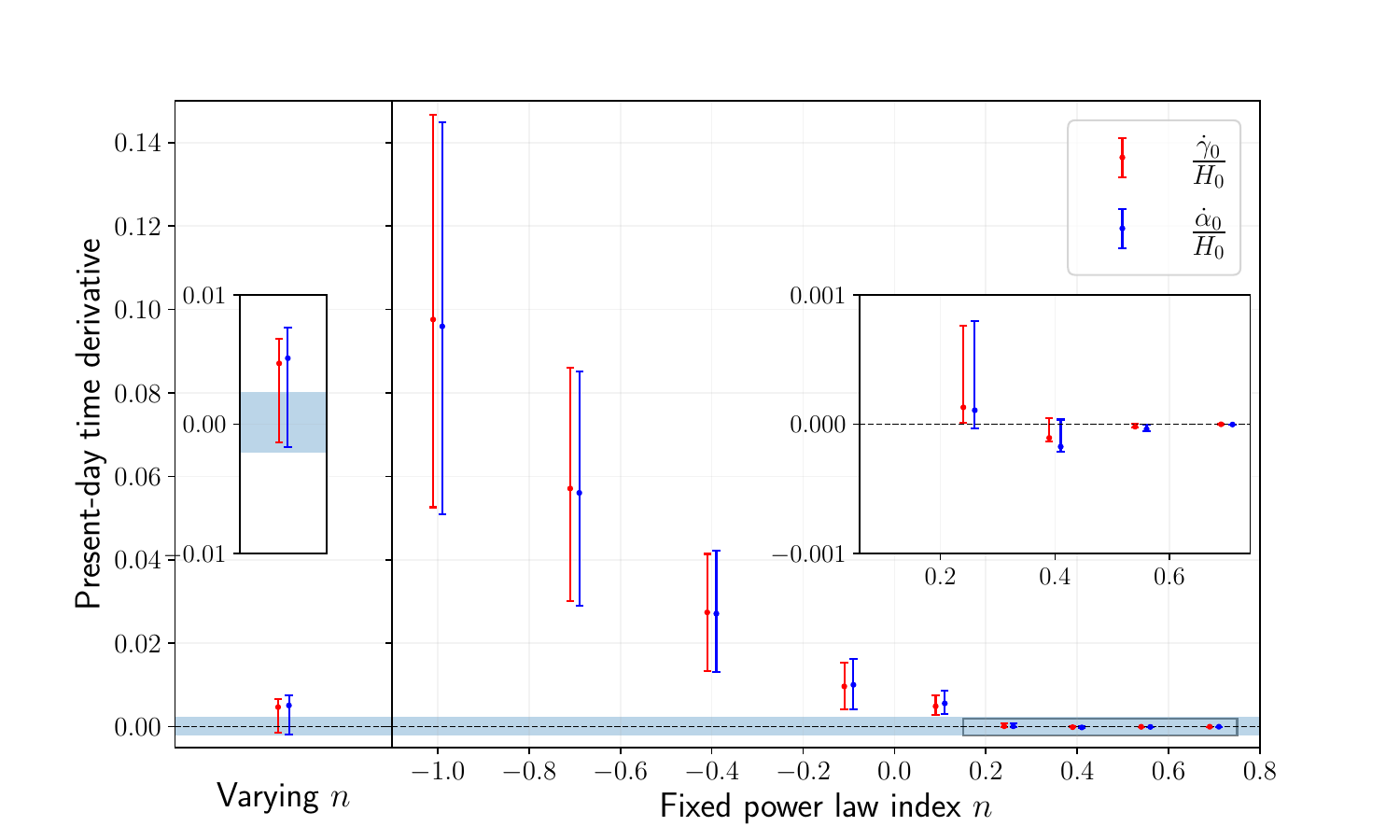}
\caption{Constraints on $\alpha'$ and $\gamma'$ at $z=0$, for different values of the fixed power-law index $n$. The shaded region is the constraint on $\dot{\alpha}$ from the ephemeris of Mars, converted to a constraint on $\alpha'$ by dividing by $H_0$. There is no equivalent constraint on $\gamma'$ from the Solar System. The power-law index is the primary determining factor of the slope of the PPN parameters at $z=0$, which is why the constraints depend strongly on $n$. Marginalising over $n$ gives constraints that are within a factor of two of the best that can be obtained from observations of the Solar System.}
\label{fig_gammaalphaprimeerror}
\end{figure}

The constraints on $\alpha'$ and $\gamma'$ at $z=0$ are shown in Fig. \ref{fig_gammaalphaprimeerror}. Due to our choice of a power-law for the time-dependence of $\alpha$ and $\gamma$, the constraints on the derivatives of these PPN functions at $z=0$ strongly depend on the value of $n$, becoming worse as the value of this power-law index decreases. The constraints on $\alpha'$ and $\gamma'$ are, however, quantitatively similar.

The constraint on ${\dot{G}}/{G}$ from the ephemeris of Mars is converted to these units by dividing by $H_0$; for this we take the Planck $\Lambda$CDM value (67.37 km s$^{-1}$Mpc$^{-1}$) and ignore any uncertainty. This results in the constraint $\alpha'=0.145^{+2.3}_{-2.3}\times 10^{-3}$ at $z=0$, which is displayed on the figure as a shaded band. For a power-law value from between $0.1$ and $0.25$ onwards, the constraints on $\alpha'$ at $z=0$ are better than those obtained within the Solar System. The constraints on $\gamma'$ at $z=0$ have no real counterpart within Solar System tests, so these should be considered as being entirely new. Interestingly, these results show that the derivatives of $\alpha$ and $\gamma$ are less well constrained for the models that most affect late-time cosmology, which leads us to expect that the inclusion of additional cosmological data (such as BAO) may improve the constraints for smaller power-laws. We will carry out this investigation in future work. We remind the reader that these are derived parameters within our framework, and that the constraints depend on the assumption of a power law form for the time evolution of the PPN parameters. As such, the constraints are not exactly equivalent to the Solar System constraints.

\subsection{Varying power-law index} \label{subsec:varying_powerlaw}

We also wish to consider cases in which the power-law is allowed to vary, and be marginalised over. In this case we need to make a choice for the upper bound of the prior on $n$. Most of our results are for the choice $n_{\rm max}=0.25$, which we will follow with a discussion of the effect of choosing larger values.

\subsubsection{Upper bound of $n_\text{max} = 0.25$}

This is the most aggressive choice of prior we consider, as it removes most of the physically uninteresting region, where the PPN parameters only contribute when standard matter is unimportant. We find that the PPN parameters $\bar{\gamma}$ and $\bar{\alpha}$ are constrained to be within $18\%$ and $21\%$ of their GR values at the $68\%$ confidence level, respectively, which is similar to the constraints where $n$ is fixed. The constraints on $n_s$, $\ln{\left(10^{10}\,A_s\right)}$, $Y_p$ and $\omega_b$ change little when the power-law is allowed to vary. This is unsurprising as these constraints were broadly constant across the fixed power-law cases, and no new degeneracies are introduced by varying the power-law.

The constraints on $H_0$, $\omega_c$ $\bar{\gamma}$ and $\bar{\alpha}$ are of order the constraints for the fixed power-law case with $n= 0.1$. For $H_0$, $\bar{\gamma}$ and $\bar{\alpha}$ this corresponds to constraints that are comparable to the worst of the fixed power-law cases, while for $\omega_c$ the constraints are better than many of the fixed power-laws (due to the increasing degeneracy for smaller power-laws, and the fact that these power-laws have little weight when marginalised over). The constraints on all four of these parameters are shown in the earlier plots, alongside the results for the fixed power-law cases. These constraints are in keeping with the behaviour seen in the fixed power-law cases, although it is encouraging that there is no strong degeneracy that makes the constraints significantly worse when the power-law is allowed to vary. This is partly due to our choice to use the average values $\bar{\gamma}$ and $\bar{\alpha}$, rather than the starting values.

\begin{figure}
    \hspace{-1cm}
    \includegraphics[width=1.1\linewidth]{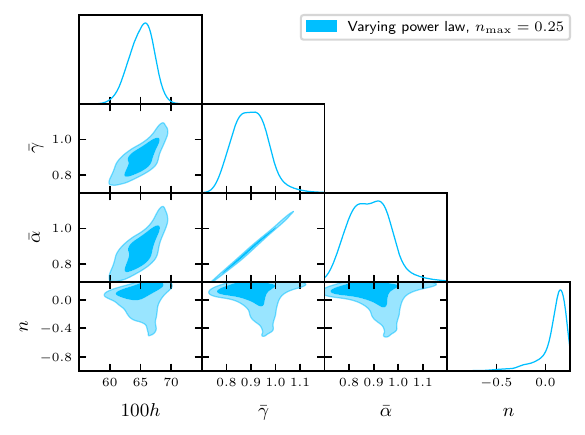}
    \caption{1D and 2D posteriors on $h$ (i.e. $H_0$), $\bar{\gamma}$\,, $\bar{\alpha}$\,, and the power law index $n$\,. For $n$\,, we have a flat prior $n \in (-15, 0.25)$, and we have applied a smoothing scale of $0.3\sigma$ to all posteriors. There is a clear change in behaviour for $n$ close to the upper bound of the prior, with a wider region that extends upwards as this upper bound is allowed to increase.}
    \label{fig_varied_powerlaw_posteriors}
\end{figure}

Some selected 2D posteriors for this case are shown in Fig. \ref{fig_varied_powerlaw_posteriors}. There remains a very strong degeneracy between the PPN parameters, as was the case for all of the fixed power law cases. The $\bar{\gamma}$ and $\bar{\alpha}$ and $H_0$ 2D degeneracies are also broadly similar to the fixed powerlaw cases. The degeneracy between each of these three parameters and the powerlaw is effectively a ``T''-shape. Around the $\Lambda$CDM values of the PPN parameters, the power law does not have a strong effect and the PPN parameters can be largely compensated for. However, once the power-law reaches the physically uninteresting region, then the roles reverse and the PPN parameters values can take a much wider range of values without significantly affecting the likelihood. This tail, where a wider range of PPN values start to be allowed, can be seen in the 2D plots, and extends further when the upper bound on the prior is increased.

The constraints on the derivatives remain of the same order as the constraints obtained for the fixed power-laws in the range $-0.1 < n < 0.1$, although they are no longer Gaussian when the power-law index is allowed to vary. The shape of the 2D posteriors between the derivatives and the power-law is not strongly meaningful, in the sense that the derivative is determined analytically by the average and the power law. The $68\%$ confidence-level constraints on the derivatives $\alpha'$ and $\gamma'$ are shown in Table \ref{table_alphadot_gammadot_vs_mars_ephemeris}, and in the varying power-law case can be seen to be worse than the Mars ephemeris constraints by a factor of less than two (though better than the Lunar Laser Ranging, Pulsars and Helioseismology results presented in Table 5 in Ref. \cite{Will_2014}). The competitiveness of this constraint is a significant result for the varying power-law case, as this analysis allows substantial freedom over the whole history of the Universe, and relies on very different assumptions to those underlying the Solar System and astrophysical constraints. 

\begin{table}
\centering
\begin{mytabular}[1.2]{|c|c|c|} 
\hline 
Parameter &  Solar System constraint & CMB constraint \\
\hline 
$\dot{\alpha}_0/H_0$ & $\left(0.15 \pm 2.3\right) \times 10^{-3}$ & $\left(5.1^{+2.4}_{-6.9}\right)\times 10^{-3}$ \\
\hline 
$\dot{\gamma}_0/H_0$ & --- & $\left(4.7^{+1.9}_{-6.1}\right)\times 10^{-3}$ \\
\hline
\end{mytabular} 
\caption{CMB constraints on ${\dot{\alpha}_0}/{H_0}$ and ${\dot{\gamma}_0}/{H_0}$, for a PPNC power-law with varying $n$\,. Solar System constraints come from the ephemeris of Mars \cite{konopliv2011mars}, which we have converted using the Planck value of $H_0$ for the best-fitting $\Lambda$CDM model. There is no equivalent Solar System constraint on $\dot{\gamma}_0\,$.}
\label{table_alphadot_gammadot_vs_mars_ephemeris}
\end{table}

\subsubsection{Increasing the upper bound on the prior}

Fig. \ref{fig_gammaalphaerror} shows a transition in the fixed power-law results between $n\approx0.2$ and $n\approx0.4$, due to the non-GR effects being pushed deeper into the radiation dominated era as $n$ increases. Because of this, we expect that increasing the upper bound for the prior on $n$ (when $n$ is allowed to vary) should result in considerable changes to the resulting constraints. In order to investigate this effect, we ran chains for prior bounds $n_\text{max}=0.4$ and $n_\text{max}=1.0$, ultimately finding that the $n_\text{max}=0.25$ case is a more sensible choice. We reached this conclusion as the larger prior bounds cause a prior volume effect due to the PPN parameters being able to take a much wider range of values without having any noticeable effect on the cosmology, which in turn results in the larger power-law values dominating the posteriors without causing any physically interesting effects. In all cases the posterior is concentrated towards the upper prior bound.

The posteriors in the $n_\text{max}=1.0$ case were strongly concentrated in the physically uninteresting region, as expected, with the result that the constraints on the PPN parameters were significantly poorer, while the constraints on the standard parameters were close to the $\Lambda$CDM case, and the degeneracies of the PPN parameters no longer exist.

The $n_\text{max}=0.4$ case was chosen as a near to $n_\text{max}=0.25$\,, but more conservative, upper bound, which arguably fully contains the transition towards the physically uninteresting region. Despite the small difference in upper prior bound, the power law posterior has a central value of around $0.26$, larger than the upper prior bound in the $n_\text{max}=0.25$ case. As a result, compared to the $n_\text{max}=0.25$ case, the degeneracies with $\omega_c$ and $H_0$ are reduced, with a corresponding improvement in the constraints on these parameters and a shift of the central value of the $H_0$ posterior back towards its value in $\Lambda$CDM. In this case the central values of the constraints on $n_s$, $\ln{\left(10^{10}\,A_s\right)}$, $Y_p$ and $\omega_b$ change little compared to the fixed power-law results, and show essentially the same behaviour as in the $n_\text{max}=0.25$ case.

With increasing $n_\text{max}$, the top of the T-shape degeneracy seen in Fig. \ref{fig_varied_powerlaw_posteriors} gets extended vertically and horizontally to larger and larger values of the PPN parameters and power-law index, with most of the posterior concentrated in the region near the upper bound on the prior for $n$. For example, the allowed ranges for $\bar{\alpha}$ and $\bar{\gamma}$ increase by more than a factor of two for $n_\text{max}=0.4$, and are weighted by the larger (more preferred) power-law values. This increase in the top of the T indicates to us that there is little point in extending the upper bound on the prior much beyond $n_\text{max}=0.25$. Interestingly, because the present-day values of $\alpha'$ and $\gamma'$ are derived parameters, which strongly depend on the power-law, their constraints become much tighter as $n_\text{max}$ increases. This is because the concentration of the posterior around larger power-law values results in a decrease in the value of these derivatives, which can cause an improvement in the constraints by more than a factor of two. However, we do not consider this to be physically meaningful, as most of the corresponding posterior would be concentrated in the physically uninteresting region.

Ultimately, we decided to use the upper prior bound of $n_\text{max}=0.25$ as it does not contain too much influence from the physically uninteresting region, which would otherwise distort our results. For $n_\text{max}=0.25$ the deviation of the PPN parameters from GR has already dropped to $\sim2\%$ of its largest value by matter-radiation equality, meaning that the deviation in the Friedmann equation from GR at this point is already reduced to no more than $1\%$ of its starting value. While $n_\text{max}=0.25$ seems like a sensible choice, we expect that there is a range of values around this value that could also be considered sensible, and for which the resulting constraints do not change significantly. It is possible that a tapered (rather than hard) prior is a better way to handle this situation; we leave this investigation to future work.

\subsection{Degeneracies and physical interpretation of constraints}

We will now explore the physical interpretation of the results obtained from our MCMC analyses, focusing on the PPN parameters $\bar{\alpha}$ and $\bar{\gamma}$, and their degeneracies with $H_0$ and $\omega_c\,$.

\subsubsection{PPNC phenomenology}
\label{sec_phenomenology}

A key aspect of the PPNC phenomenology is that the cosmological behaviour is substantially closer to $\Lambda$CDM when $\bar{\alpha}=\bar{\gamma}$. There are two interlinked causes of this: (i) the novel term $\mathcal{G}\mathcal{H}\Psi$ in the momentum constraint (\ref{eq_pert3}), and (ii) a cancellation that occurs in the PPNC phenomenology when $\bar{\alpha}\approx\bar{\gamma}$. We will discuss each of these in turn, below.

\begin{figure}
    \centering
    \begin{subfigure}{0.95\textwidth}
    \includegraphics[width=\linewidth]{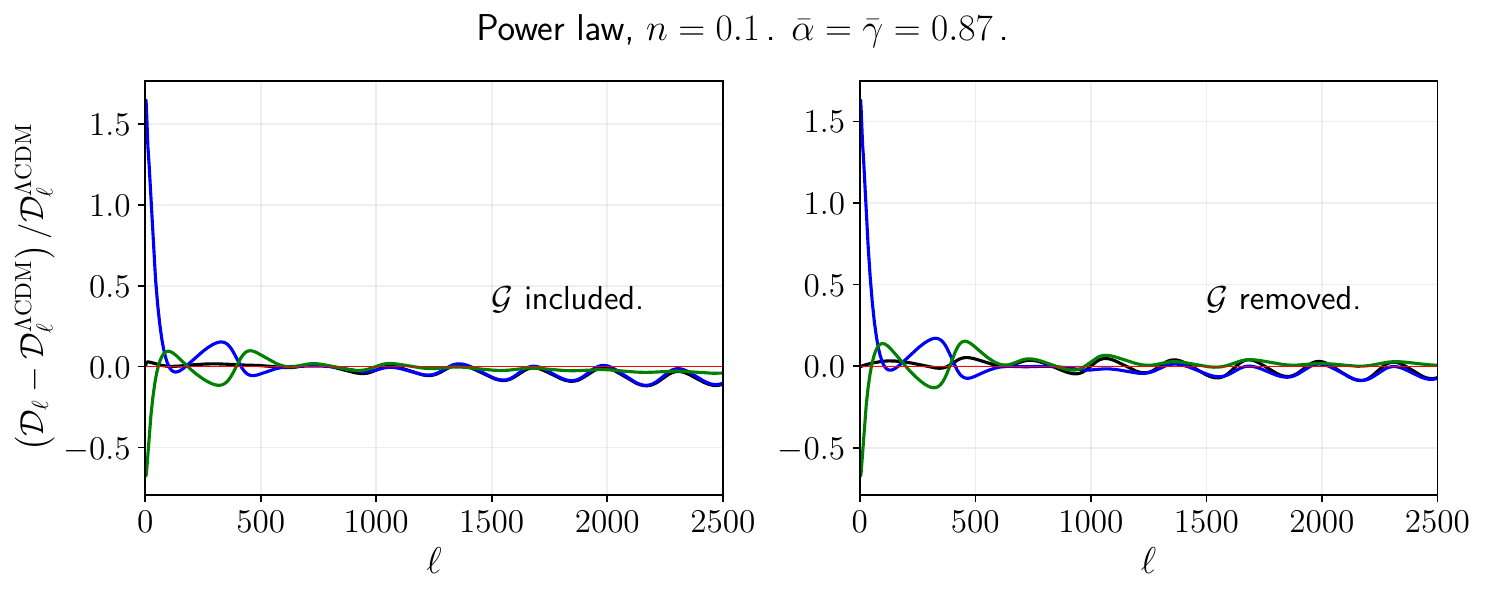}
    \end{subfigure}
    \vskip -1ex
    \begin{subfigure}{0.95\textwidth}   \includegraphics[width=\linewidth]{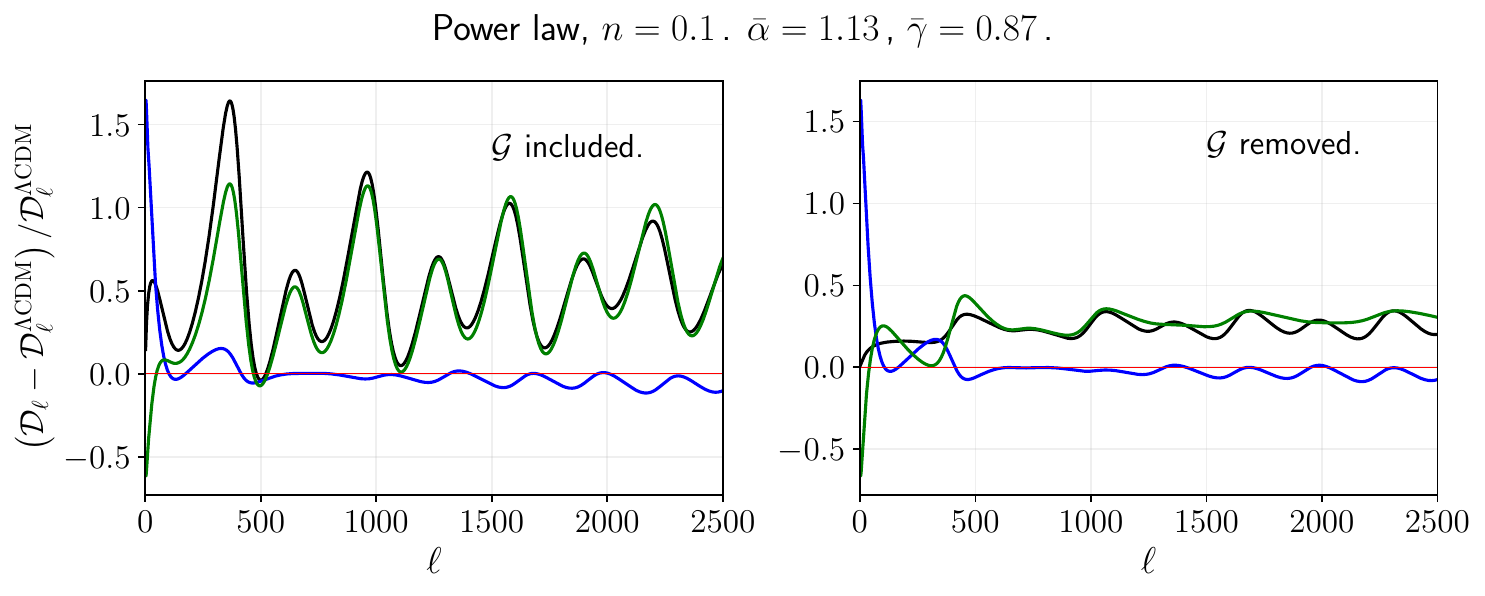}
    \end{subfigure}
    \vskip -2ex
    \caption{The relative effects of PPNC background and perturbation evolution on $\mathcal{D}_{\ell}^{TT}\,$. In each case, we first use the full PPNC equations, then re-calculate with the $\mathcal{G}$ function artificially removed. Results are shown for a power law with $n = 0.1$. Top plots have $\bar{\alpha} = \bar{\gamma} = 0.87\,$, and bottom plots have $\bar{\alpha} = 1.13$ and $\bar{\gamma} = 0.87\,$. Red curves show $\Lambda$CDM, black full PPNC, green when PPNC perturbation equations are used but the background expansion is $\Lambda$CDM, and blue when the background is evolved using the PPNC Friedmann equation but perturbations evolve under $\Lambda$CDM equations. It can be seen that $\bar{\alpha} = \bar{\gamma}$ gives more similar phenomenology to $\Lambda$CDM, and removing the $\mathcal{G}$ function removes most of the difference created when $\bar{\alpha} \neq \bar{\gamma}$. }
\label{fig_background_vs_perturbations}
\end{figure}

\textit{(i) Novel term}. In the $\Lambda$CDM limit, the $\mathcal{G}\mathcal{H}\Psi$ term in the momentum constraint equation (\ref{eq_pert3}) does not exist. For sub-horizon perturbations in the PPNC case, $\mathcal{G}$ is given by $\left(\alpha - \gamma\right)/\gamma + \hat{\gamma}/\gamma\,$, so that when $\bar{\alpha} \neq \bar{\gamma}$ the term $\mathcal{G}\mathcal{H}\Psi$ drives the evolution of $\Psi$. This produces distinct phenomenology in the CMB, which is very different to $\Lambda$CDM. This new phenomenology is shown\footnote{It is similarly shown for the Doppler and Sachs-Wolfe source terms in Fig. \ref{fig_background_vs_perturbations_2}. These are the most important source terms when considering the effect of the PPN parameters on the CMB.} in Fig. \ref{fig_background_vs_perturbations} 
, which presents the effects on $\mathcal{C}_{\ell}^{TT}$ of the PPNC modifications with and without the $\mathcal{G}$ term. The green lines display the effect of the PPNC modifications to the perturbation equations, with the key point that they differ substantially from $\Lambda$CDM-like behaviour for all cases except $\bar{\alpha} = \bar{\gamma} = 0.87$. In the case where $\left(\bar{\alpha}, \bar{\gamma}\right) = \left(1.13, 0.87\right)$, it can be seen that the difference in the evolution of the perturbations, compared to $\Lambda$CDM, is much reduced by artificially removing the $\mathcal{G}$ term from the modified perturbation equations.

\begin{figure}[h]
\begin{subfigure}{0.52\textwidth}
\hspace{-2.6cm}
    \includegraphics[width=1.25\linewidth]{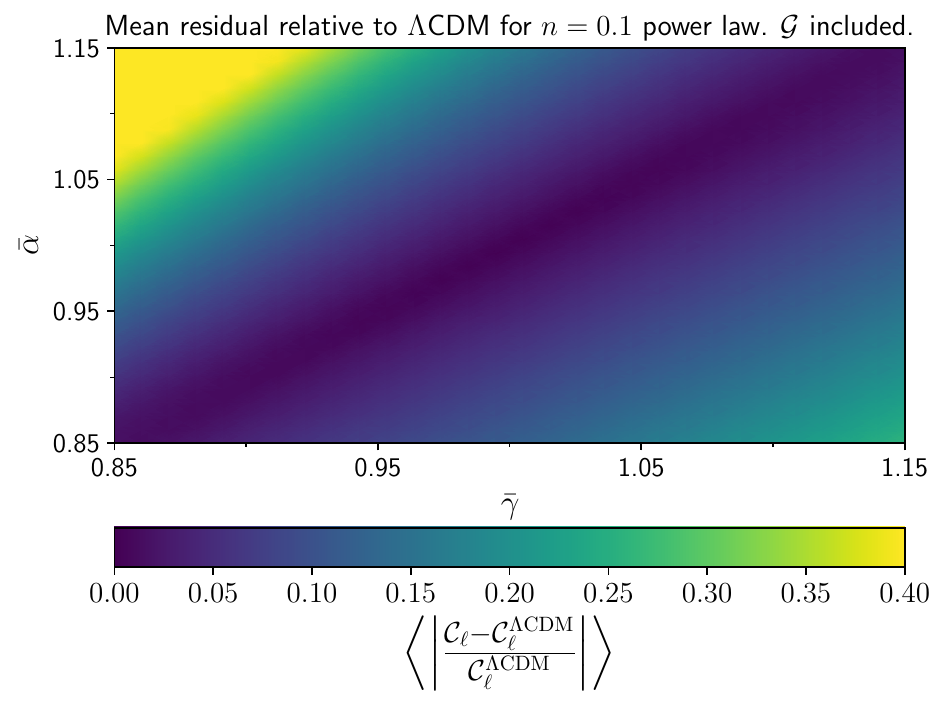}
    \caption{}
    \label{fig_mean_residual_0pt1_with_G}
\end{subfigure}
\hspace{-0.7cm}
\begin{subfigure}{0.52\textwidth}
    \includegraphics[width=1.25\linewidth]{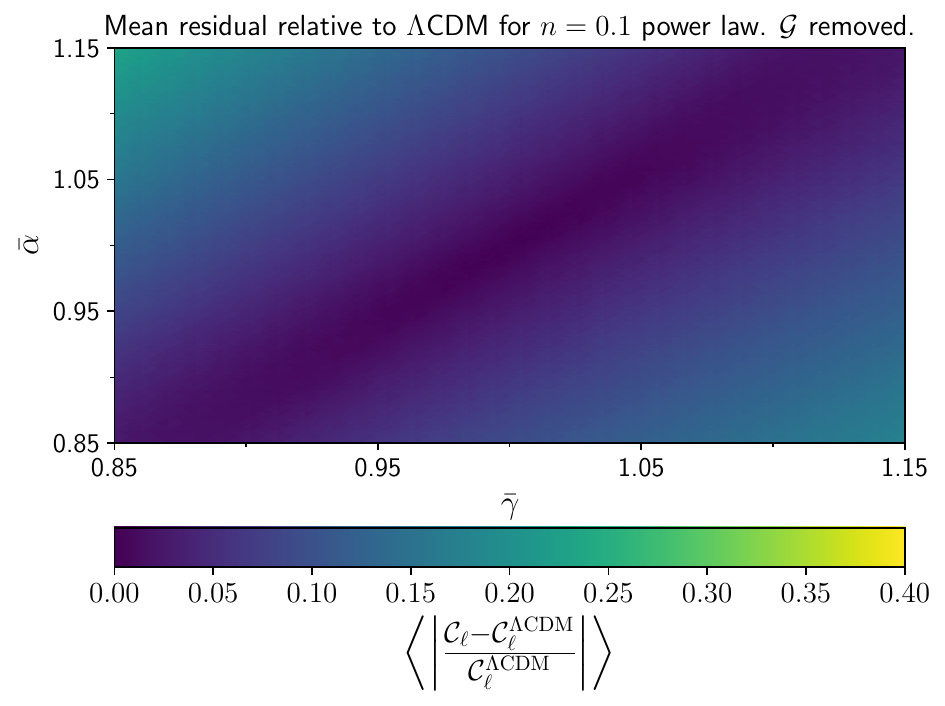}
    \caption{}
    \label{fig_mean_residual_0pt1_no_G}
\end{subfigure}
\caption{Colour maps of the mean absolute residual of $\mathcal{C}_{\ell}$ relative to $\Lambda$CDM, averaged over multipoles $\ell \in \left[2,1001\right]$. Results are displayed as a function of $\bar{\alpha}$ and $\bar{\gamma}$, for the $n = 0.1$ power law. In the left panel, we have used the full set of PPNC equations. In the right panel, the term $\mathcal{G}\mathcal{H}\Psi$ has been artificially removed from the momentum constraint equation, showing that $\mathcal{G}$ drives the preference for $\bar{\alpha} \approx \bar{\gamma}$\,.}
\label{fig_mean_residual_0pt1}
\end{figure}

The $\mathcal{G}$ term generally affects all potentials in the sub-horizon regime, with a roughly uniform amplification of the Weyl potential $\Phi + \Psi$ across all scales with $k \gtrsim k_H$, when $\bar{\gamma} < \bar{\alpha}\,$. This behaviour can be explained by combining the $\mathcal{G}$ term with the existing damping term, such that the total damping is given by $\mathcal{H}\left(\Phi - \mathcal{G}\Psi\right)$. Combined with the slip relation on small scales, $\Phi = ({\alpha}/{\gamma})\,\Psi$, this can be written as $\mathcal{H}\Phi\left(1 - \mathcal{G}\,{\gamma}/{\alpha}\right)$, so when $\mathcal{G}$ is positive ($\bar{\gamma} < \bar{\alpha}\,$), the $\mathcal{G}$ term strongly reduces the damping of the potentials, and thus allows them to grow much more rapidly than in $\Lambda$CDM. This conclusion holds across the PPNC parameter space, as shown in Fig. \ref{fig_mean_residual_0pt1_with_G}, which shows the mean absolute residual in $\mathcal{C}_{\ell}^{TT}$  (averaged over $\ell \in \left[2, 1001\right]$) relative to the best-fit $\Lambda$CDM, for a PPNC power law with $n = 0.1$\,. This mean absolute residual is  $80.2\,\%$ for $\left(\bar{\alpha}, \bar{\gamma}\right) = \left(1.15, 0.85\right)$, whereas it is $1.67\,\%$ for $\left(\bar{\alpha}, \bar{\gamma}\right) = \left(0.85, 0.85\right)$, and $1.61\,\%$ for $\left(1.15, 1.15\right)$\,, even though in all of these cases $\vert \bar{\alpha} -1 \vert$ and $\vert \bar{\gamma} - 1 \vert$ are the same. 

\textit{(ii) Background \& perturbation cancellation}. When $\alpha = \gamma$,\footnote{Note that $\bar{\alpha}$ and $\bar{\gamma}$ being equal means that $\alpha(a) = \gamma(a)$ for all $a$, as they have the same power-law index $n$, and we are implementing $\gamma = \alpha= 1$ at $a=1$.} the PPNC modifications to the background and perturbation equations have roughly equal and opposite effects on the gravitational potentials at last scattering, resulting in an effective cancellation, and thus a return to behaviour similar to the $\Lambda$CDM case. For $\gamma > 1$, this occurs due to the coupling parameter $\mu$ from Eqns. (\ref{eq_pert1}) and (\ref{eq_pert3}) becoming larger, and enhancing the growth of perturbations is enhanced, while at the same time increasing the expansion rate $H$ from Eq. (\ref{eqn_Friedmann_code}), which suppresses the growth of perturbations. These two effects are approximately equal in size, and so cancel each other.

\begin{figure}
    \hspace{-1cm}
    \begin{subfigure}{1.1\textwidth}
    \includegraphics[width=\linewidth]{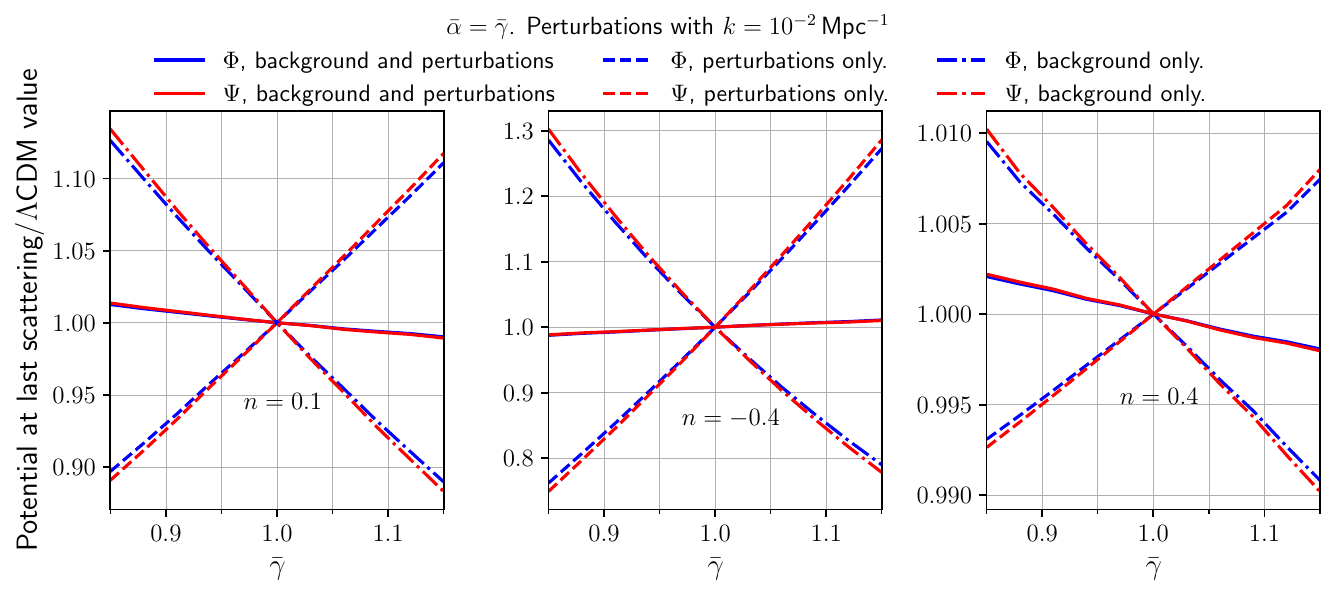}
    \end{subfigure}
    \vskip -1ex
    \hspace{-1cm}
    \begin{subfigure}{1.1\textwidth}
    \includegraphics[width=\linewidth]{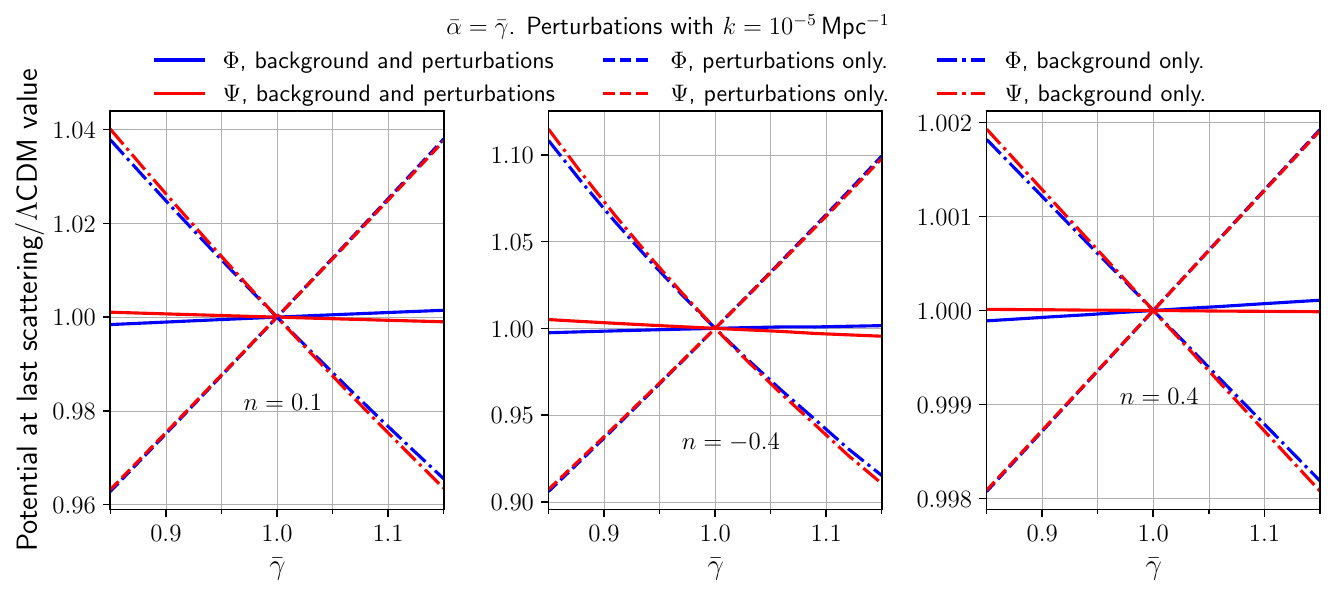}
    \end{subfigure}
    \caption{Metric perturbations $\Phi$ and $\Psi$ at $a_{\rm LS}$, as a function of $\bar{\gamma}$, for a subhorizon mode with $k = 10^{-2}\, {\rm Mpc}^{-1}$ (top) and a superhorizon mode with $k = 10^{-5}\,{\rm Mpc}^{-1}\,$ (bottom). They are evaluated for the case where the full PPNC equations are used, for the case where only the perturbations are governed by PPNC equations (and the background by $\Lambda$CDM equations), and for the case where only the background is governed by PPNC equations (and the perturbations by $\Lambda$CDM). We have set $\bar{\alpha} = \bar{\gamma}$, and normalised each perturbation by its $\Lambda$CDM value. Results are shown for the $n = 0.1$\,, $-0.4$ and $0.4$ power laws, but the cancellation between the effects of the modified background and perturbations shown here persists for a wide range of power laws and wavenumbers.}
    \label{fig_Phi_Psi_LS}
\end{figure}

This cancellation effect is shown graphically in Fig. \ref{fig_Phi_Psi_LS}, where the values of $\Phi$ and $\Psi$ at last scattering are shown as a function of $\bar{\gamma} = \bar{\alpha}$ for the $n = 0.1$, $-0.4$ and $0.4$ power laws. The different lines show the background-only, perturbation-only and combined effects on the potentials, and it can be seen that the effects of the background and perturbations are opposite and approximately equal for a wide range of values for $\bar{\gamma}$, for a wide range of values of $n$, and for both potentials. The two sets of panels show that the same cancellation mechanism applies over a wide range of $k$-modes, from deep inside the horizon ($k = 10^{-2}\,{\rm Mpc}^{-1}$) to far outside it ($k = 10^{-5}\,{\rm Mpc}^{-1}$). We also examined several values of $k$ in between these two values and found the same behaviour.

 Unsurprisingly, the cancellation of the PPNC effects in the metric potentials carries over to the Sachs-Wolfe and Doppler source terms, and thus to the CMB spectra, as the Sachs-Wolfe term is almost entirely determined by $\Phi_{\rm LS}$, while the Doppler term is also determined by it through the Euler equation evaluated at $a_{\rm LS}\,$. We demonstrate this in Fig. \ref{fig_background_vs_perturbations} by showing the effects on $\mathcal{C}_{\ell}^{TT}$ of the PPNC modifications to the background (blue lines) and perturbation (green lines) equations separately,
 \footnote{i.e. we evolve the background with the PPNC equations and the perturbations with the $\Lambda$CDM equations, and vice versa.} and combined (black lines). Both the PPNC background-only and PPNC perturbations-only cases can individually display large residuals of $\mathcal{C}_{\ell}$ from $\Lambda$CDM. In the case $\bar{\alpha} = \bar{\gamma} = 0.87$, with an $n = 0.1$ power law, these effects cancel well, and the power spectrum for the full PPNC system (background and perturbations) remains close to $\Lambda$CDM for all multipoles. The lower row of panels show that in the case $\left(\bar{\alpha}, \bar{\gamma}\right) = \left(1.13, 0.87\right)$, there is virtually no cancellation.  Instead, the effects of the background and perturbations combine constructively to further amplify the acoustic peaks, while the $\ell \gtrsim 500$ region is dominated by the PPNC perturbations, which deviate strongly from $\Lambda$CDM. We see that the absence of a cancellation in the case $\bar{\alpha} \neq \bar{\gamma}$ is much less severe (but still present at a lower level) if we artificially remove the $\mathcal{G}$ term from the PPNC momentum constraint equation, as shown in the right half of the figure.

Combining results (i) and (ii) shows that $\bar{\alpha} \approx \bar{\gamma}$ is required to prevent the $\mathcal{G}$ term from generating substantial non-$\Lambda$CDM-like behaviour in the perturbations, and that when $\bar{\alpha} \approx \bar{\gamma}$ the effects of the PPN parameters on the metric potentials largely cancel due to competing changes to the background and perturbation equations. As this cancellation removes the PPNC effects on the metric perturbations, when $\bar{\gamma}\approx \bar{\alpha}$ the remaining effect of the PPN parameters is on background quantities. The most significant differences in the CMB power spectrum between the PPNC and $\Lambda$CDM cases are that the evolving $\gamma(a)$ modifies the angular diameter distance to last scattering and the sound horizon at last scattering, which generically causes a shift in the peak locations through the ratio $\theta(a_{\rm LS}) = r_s(a_{\rm LS})/d_A(a_{\rm LS})\,$ (unless the changes to $d_A(a_{\rm LS})$ and $r_s(a_{\rm LS})$ conspire to cancel, which does not typically occur).

\begin{figure}
    \hspace{-2.5cm}
    \includegraphics[width=1.25 \linewidth]{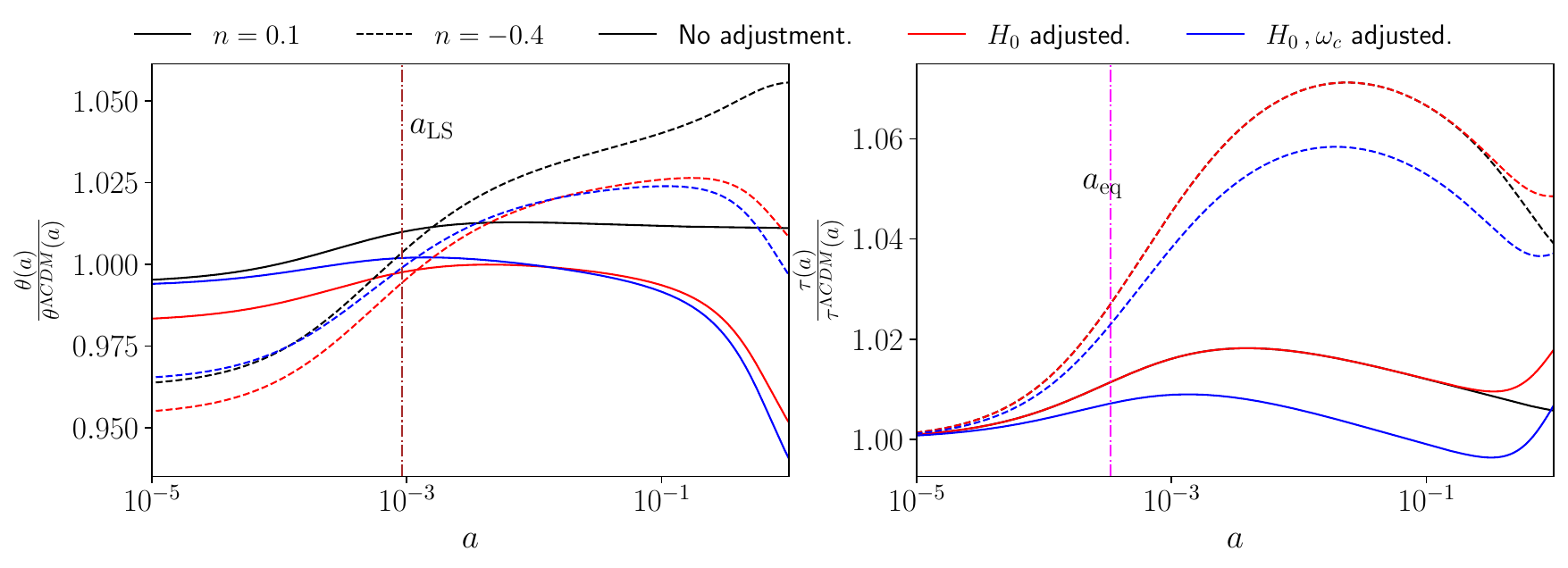}
    \caption{Comparison of the effect on the cosmological background expansion of varying $\bar{\gamma}$ (i.e. changing gravity) while holding all the standard $\Lambda$CDM parameters fixed (black), to the effect of varying $\bar{\gamma}$ by the same amount but adjusting $H_0$ (red), or both $H_0$ and $\omega_c$ (blue), to compensate its effect on the FLRW expansion.
    Results are displayed for the $n = 0.1$ (solid) and $n = -0.4$ (dashed) cases, with $\bar{\gamma} = 0.85$ so that the curves $\gamma(a)$ for each power law are as given in Fig. \ref{fig_gamma_of_a_power_laws}. In these plots $\theta(a) = r_s(a)/d_A(a)\,$, and all curves have been normalised by their $\Lambda$CDM evolution. The quantities $\theta(a_{\rm LS})$ and $\tau(a_{\rm eq})$ respectively determine the locations and heights of the CMB acoustic peaks. These plots show that $H_0$ and $\omega_c$ can be used to compensate the effects of the PPN parameters on the background expansion.}
    \label{fig_background_gamma_H0_omc}
\end{figure}

The effect of the PPNC power-law on the cosmological background is demonstrated in Fig. \ref{fig_background_gamma_H0_omc}. The left-hand plot shows $\theta(a)$ relative to $\Lambda$CDM\,. The black curves show the effect of varying $\bar{\gamma}$ while holding all the standard cosmological parameters fixed, whereas the red curves show the effect of varying $\bar{\gamma}$ while also compensating its effect by varying $H_0\,$. Without adjusting $H_0$, setting $\bar{\gamma} < 1$ shifts $\theta(a_{\rm LS})$ above its $\Lambda$CDM value, which would correspond to a lower $l$ for the first acoustic peak. However, it can be compensated by reducing $H_0$ relative to $\Lambda$CDM, so that the peak locations return to their observed values. As such we expect there to be a positive degeneracy between $\bar{\gamma}$ and $H_0$ (when $\bar{\alpha}\approx\bar{\gamma}$). The right-hand plot in Fig. \ref{fig_background_gamma_H0_omc} shows the conformal time $\tau$ that has elapsed at a given $a$ since the beginning of the PPNC evolution at $a_1 = 10^{-10}\,$, as a ratio of its $\Lambda$CDM value. Of particular importance is $\tau_{\rm eq} = \tau(a_{\rm eq})\,$. Setting $\bar{\gamma} < 1$ increases $\tau_{\rm eq}\,$, because weakening gravity at the level of the background means that the effective energy density of cold dark matter, $\gamma \rho_c\,$, which enters into the Friedmann equation (\ref{eqn_Friedmann_code}), is always lower than in $\Lambda$CDM. Thus, the radiation era lasts longer in $\tau$, and so acoustic oscillations in the photon-baryon fluid have more time to drive acoustic anisotropies in the CMB before dark matter takes over and suppresses the oscillations. Such an amplification of the acoustic peaks from a reduction in $\bar{\gamma}$ can also be created by lowering $\omega_c\,$. Hence, the effect on $\tau_{\rm eq}$ of reducing $\bar{\gamma}$ from one can be compensated, at least in part, by a suitable increase in $\omega_c\,$, as shown by the blue curves, which indicate a negative degeneracy between $\bar{\gamma}$ and $\omega_c$ (when $\bar{\alpha}\approx\bar{\gamma}$).

We have described the phenomenology of the PPNC modifications in terms of the importance of the $\mathcal{G}$ term, the cancellation between the background and perturbation effects on the metric potentials, and the remaining effect of the PPN parameters on the background quantities. This description broadly holds over most of the range of powerlaws considered in this work, with the exception of powerlaws with $n \gtrsim 0.4$ (deep into the physically uninteresting region). In this case virtually all the evolution in $\alpha(a)$ and $\gamma(a)$ happens for $a \ll a_{\rm LS}$, so by the time of matter-radiation equality $\alpha$ and $\gamma$ are close to unity\footnote{Within $0.3\%$ of one for $n=0.4$ and $\left(\bar{\alpha}, \bar{\gamma}\right) = \left(0.87, 0.87\right)$.}, and hence there is no noticeable effect from $\bar{\alpha} \neq 1$ or $\bar{\gamma} \neq 1$. Removing $\mathcal{G}$ has no effect in this case.

\subsubsection{Interpretation of results}
\label{sec_interpretation}

The phenomenology described above drives the constraints, preferences and degeneracies that are seen in the data, as follows:

\textit{PPNC degeneracy.} One of the most notable features of the posteriors obtained in our MCMC analyses is that in each case $\bar{\alpha}$ and $\bar{\gamma}$ are strongly degenerate and there is a strong preference for $\bar{\alpha}$ and $\bar{\gamma}$ to be roughly equal. In general, the difference $\vert \bar{\alpha} -\bar{\gamma} \vert$ in the posteriors is constrained to be much closer to zero than either $\vert \bar{\alpha} - 1 \vert$ or $\vert \bar{\gamma} - 1 \vert$, as shown in Figs. \ref{fig_h0omcerror2d} and \ref{fig_varied_powerlaw_posteriors}. This degeneracy is caused by the phenomenology of the $\mathcal{G}$ term discussed above, and as expected it does not change significantly for power-laws with $ n \leq 0.25$ (and also holds for larger power laws).

\begin{figure}
    \centering
    \vspace{-2cm}
    \begin{subfigure}{0.85\textwidth}
    \includegraphics[width=\linewidth]{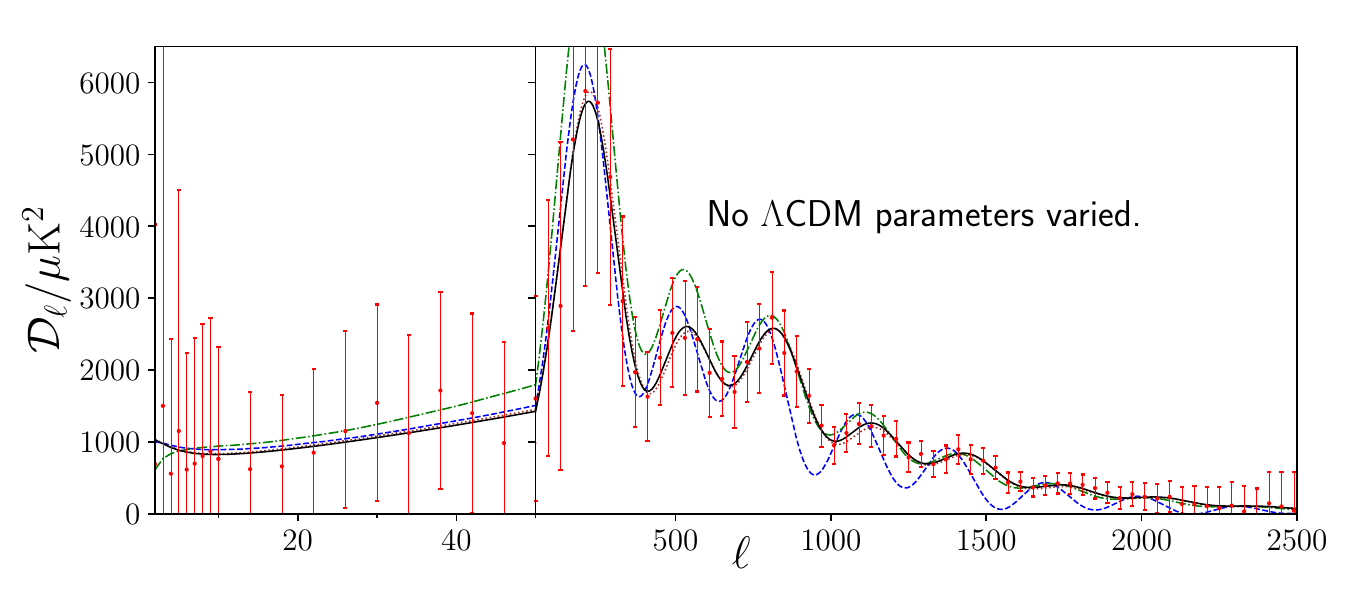}
    \end{subfigure}
    \vskip -1ex
    \vspace{-0.65cm}
    \begin{subfigure}{0.85\textwidth}
    \includegraphics[width=\linewidth]{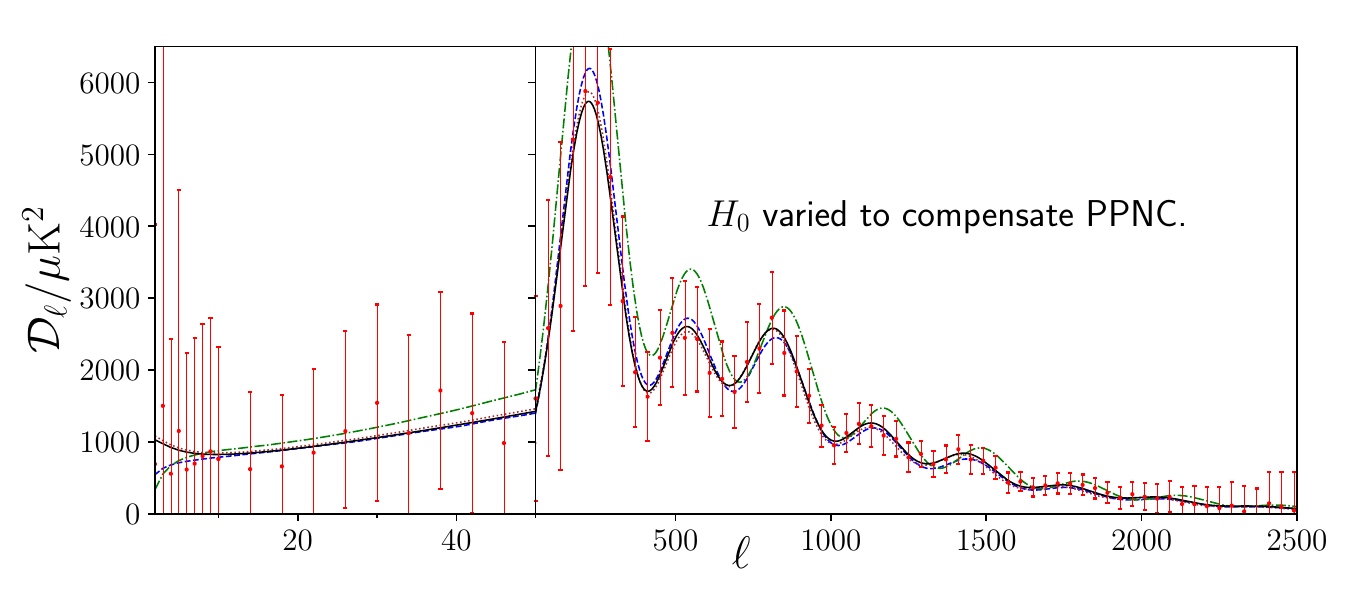}
    \end{subfigure}
    \vskip -1ex
    \vspace{-0.65cm}
    \begin{subfigure}{0.85\textwidth}
    \includegraphics[width=\linewidth]{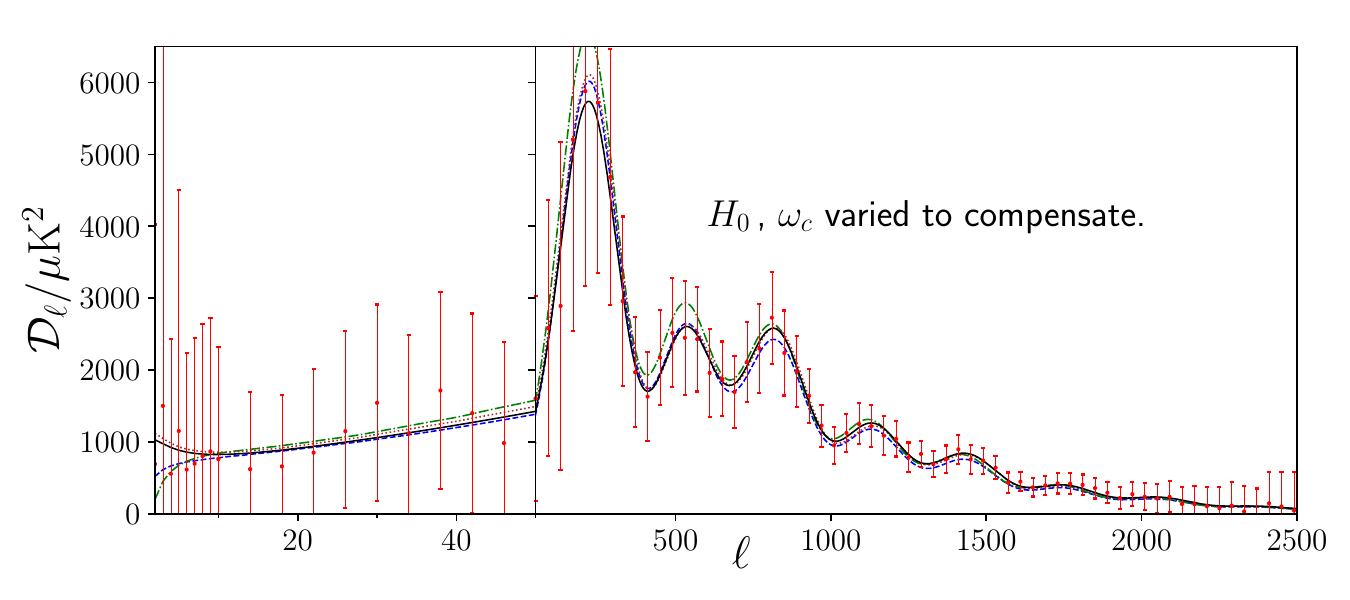}
    \end{subfigure}
    \vskip -1ex
    \vspace{-0.65cm}
    \begin{subfigure}{0.85\textwidth}
    \includegraphics[width=\linewidth]{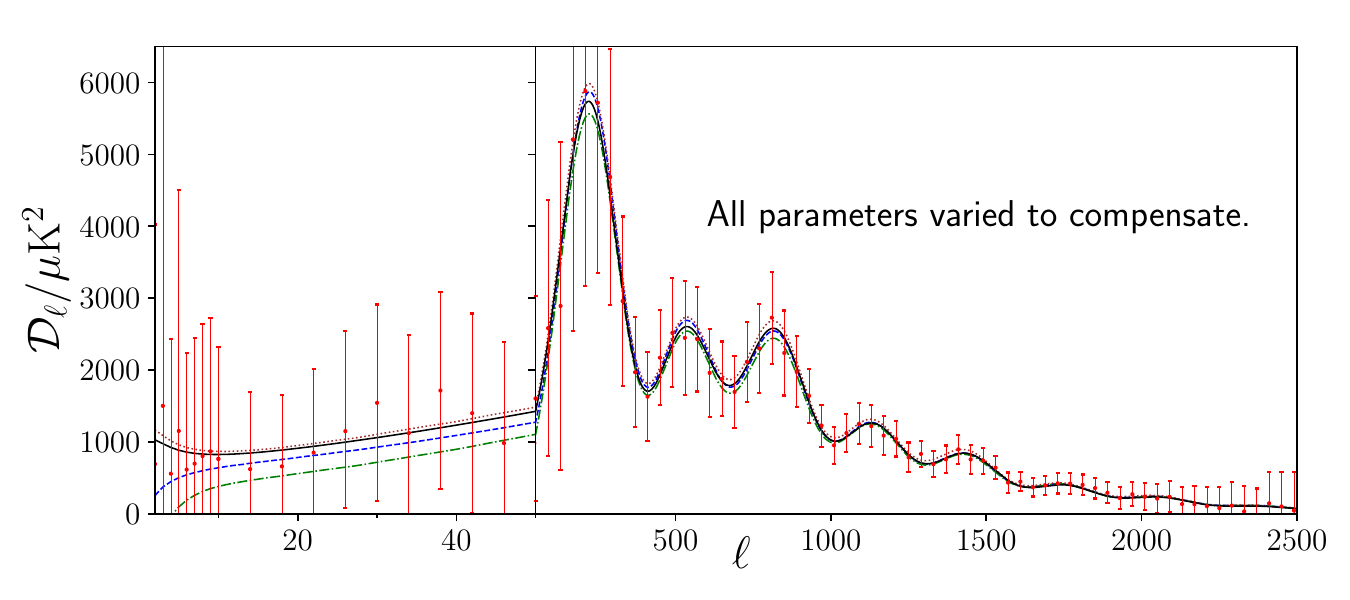}
    \end{subfigure}
    \vspace{-0.25cm}
    \caption{The angular power spectrum for $n = 0.1$ (blue), $-0.4$ (green) and $0.4$ (brown), with their best-fit values of $\bar{\alpha}$ and $\bar{\gamma}$. Planck 2018 data points are shown in red, with error bars enlarged by a multiple of five, and the corresponding best-fit $\Lambda$CDM curve is in black. In the first plot, all the basic $\Lambda$CDM cosmological parameters have been fixed. 
    In the second plot, $H_0$ has been adjusted to its best-fit value for each power-law. In the third, both $H_0$ and $\omega_c$ have been adjusted,
    and in the fourth all parameters have been adjusted. Curves have been shifted by $\mathcal{D}_{\ell}^{\rm PPNC} \rightarrow \mathcal{D}_{\ell}^{\Lambda {\rm CDM}} + 10\left(\mathcal{D}_{\ell}^{\rm PPNC} - \mathcal{D}_{\ell}^{\Lambda {\rm CDM}}\right)$ to make deviations from $\Lambda$CDM visible. Adjusting the $\Lambda$CDM parameters to compensate the effects of the PPN parameters allows for the peak heights and locations to be made very similar to $\Lambda$CDM.}
    \label{fig_Cl_1_adjust_parameters}
\end{figure}

\textit{Degeneracy with $\Lambda$CDM parameters.} The degeneracy between the PPN parameters, as discussed in detail above, means that the degeneracies between $\bar{\gamma}$ and $H_0$, and between $\bar{\gamma}$ and $\omega_c$, give rise to roughly the same degeneracies between $\bar{\alpha}$ and the $\Lambda$CDM parameters. The most significant degeneracy is between $\bar{\gamma}$ and $H_0$. The next-strongest degeneracy is between $\bar{\gamma}$ and $\omega_c$\,. Fig. \ref{fig_Cl_1_adjust_parameters} shows $\mathcal{C}_{\ell}^{\rm TT}$ for the $n = -0.4$, $n = 0.1$ and $n = 0.4$ power-laws, where $\bar{\alpha}$ and $\bar{\gamma}$ are set by their best-fit values. We first adjust $H_0$ only, then both $H_0$ and $\omega_c$, then all the rest of the $\Lambda$CDM parameters, to their best-fit values from the MCMC.

The change induced by allowing $H_0$ to be adjusted shows how the effect of the PPN parameters on the locations of the acoustic peaks can be compensated, as expected from the phenomenology noted in Section \ref{sec_phenomenology}. Increasing $H_0$ has the opposite effect to increasing $\gamma$, which results in a positive degeneracy between $\bar{\gamma}$ and $H_0$. As our fixed power-law analyses tend to indicate a preference for $\bar{\gamma} < 1\,$, they therefore push $H_0$ to lower values than the $\Lambda$CDM Planck best-fit, meaning that the Hubble tension is worsened in the PPNC approach. The next-strongest degeneracy is the negative one between $\bar{\gamma}$ and $\omega_c$.  Fig. \ref{fig_Cl_1_adjust_parameters} shows that the main effect of adjusting $\omega_c$ is to compensate the effect of the evolution of $\gamma$ on the heights of the acoustic peaks, as expected from the phenomenology noted in Section \ref{sec_phenomenology}. The effect of the PPN parameters on the background quantities, and resulting degeneracies, means that the degeneracy between $H_0$ and $\omega_c$ also changes slightly.

The degeneracy between $\bar{\gamma}$ and $H_0$ does not change significantly between the different powerlaws (until the power law starts to enter the physically uninteresting region), as seen in Fig. \ref{fig_h0omcerror2d}. The degeneracy with $\omega_c$ does change, because as the powerlaw decreases, a larger change to $\omega_c$ is required to compensate the change caused by $\bar{\gamma}$. For both $\Lambda$CDM parameters, as the physically uninteresting region is reached the degeneracies with the PPN parameters start to weaken. The behaviour of the degeneracies is also shown in Fig. \ref{fig_h0omcerror}, where the marginalised constraints on $\omega_c$ get worse for smaller $n$ due to the degeneracy getting stronger, while the marginalised constraints on both $\omega_c$ and $H_0$ get better for large $n$.

\subsection{Advantages over standard PPN tests}

From the point of view of precision tests of gravity, there are several advantages to the PPNC approach. It allows tests of gravity to be extended outside of the Solar System and Milky Way, and to be extended to new observables and more data, including very different environments and contexts with different assumptions and modelling. This not only increases the robustness of the resulting constraints, but also allows for combined and coherent tests of gravity across all astrophysical and cosmological times and spatial scales, and for cosmological modelling to be automatically consistent with existing constraints on smaller scales. The other key advantage is the ability to learn about the time evolution of the PPN parameters, which is in some ways either poorly constrained by astrophysical and Solar System tests, or not constrained at all.

Unpacking the time variation in a bit more detail, standard PPN tests typically constrain $\gamma$ at the present time, but make no statement about $\bar{\gamma}$. Solar System tests also constrain $\alpha'$ at the present time, which can be extrapolated to a constraint on $\bar{\alpha}$ only if no freedom, uncertainty or marginalisation is allowed in the functional form of the time dependence, which is a very strong assumption. The extension to cosmology laid out in this work considers constraints in two cases: those in which the power-law behaviour of the PPN parameters is fixed or allowed to vary. The fixed power-law case gives constraints on $\bar{\gamma}$ and $\gamma'$ at $a=1$, which have no equivalent in the Solar System tests of gravity. This case also gives a constraint on $\bar{\alpha}$ that is broadly comparable to standard PPN tests, being better or worse depending on which power law and test is being compared. The cosmological constraints also have the advantage of being an integral over time, so they have more support over a wider range of times and will therefore smooth out more complicated behaviour, making the power-law assumption less restrictive than in the case of Solar System tests. 

The varying power-law case has even more advantages, in that the present day rate of change of the parameters is decoupled from their values at earlier times. In particular, this means that the constraint on $\bar{\alpha}$ is no longer comparable with na{\" i}ve extrapolations from present day rate of change measurements. As stated above, this case also gives a present-day constraint on $\alpha'$ that is again broadly comparable with Solar System tests. As the varying power-law case has the weakest assumptions about the behaviour of the parameters over time, we consider this case to be the most important of the constraints obtained in this work, and the one that is most complementary to local astrophysical tests of gravity. Finally we note that as the data strongly prefer $\bar{\alpha}\simeq \bar{\gamma}$, observers performing local tests of gravity at different times would typically find local gravitational physics to be compatible with GR (with different values for $G$), despite considerable variation of $\alpha$ and $\gamma$ being allowed over cosmic history. This further reinforces the importance of combining cosmological tests with local tests.

\section{Conclusion}

We have presented the first cosmological constraints on the PPN parameters. Specifically, we have used the CMB to constrain the average values of ${\alpha}$ and ${\gamma}$ over cosmic history. We do this both for power-law evolution with fixed indices, and also by marginalising over the power-law index. In all cases we consider, the PPN parameters are constrained to be within $30\%$ of their GR values at the $95\%$ confidence level. This study lays the groundwork for unified precision tests of gravity across all astrophysical and cosmological length and time scales, using a single set of parameters. The key findings from our study are as follows:
\begin{itemize}
    \item CMB data strongly prefer $\bar{\alpha}=\bar{\gamma}$ in all cases we have studied, mirroring Solar System constraints at $z=0$. 
    \item We find consistency with GR at the $95\%$ confidence level for all fixed power-laws with $-1\leq n\leq0.25$ (and for varying $n$ with the same upper limit), except $\bar{\alpha}$ for $n=0.1$ which is consistent at the $99.7\%$ confidence level.
    \item When marginalising over $n$ we find $\bar{\alpha}=0.89^{+0.08}_{-0.09}$ and $\bar{\gamma}=0.90^{+0.07}_{-0.08}$ at the $68\%$ confidence level, which are compatible with GR at the $95\%$ confidence level. Constraints obtained with this level of freedom in the time evolution have no Solar System or astrophysical equivalent.
    \item Marginalising over $n$ gives $\alpha'=\left(5.1^{+2.4}_{-6.9}\right)\times 10^{-3}$ and $\gamma'=\left(4.7^{+1.9}_{-6.1}\right)\times 10^{-3}$ at the present time and at $68\%$ confidence level. The first of these is within a factor of two of the best Solar System constraints, while the second has no Solar System equivalent.
    \item For the fixed power-law cases in the range  $n \in [-1, 0.25]$, the $68\%$ confidence intervals for $\bar{\alpha}$ and $\bar{\gamma}$ vary between $0.04$ and $0.16$ (with little difference between the two), and have the parameters being constrained to be within $4$ to $25\%$ of their GR values.
    \item There is a strong degeneracy between $\bar{\alpha}$ and $\bar{\gamma}$, due to (i) the novel term containing $\mathcal{G}$ in the momentum constraint equation, and (ii) a cancellation between the effects of the background and perturbation equations on the behaviour of the potentials when $\bar{\gamma}\approx\bar{\alpha}$.
   \item Differences between $\Lambda$CDM and PPNC cosmologies with $\bar{\gamma}\approx\bar{\alpha}$ are primarily due to the background, and can thus be largely compensated for by adjusting $H_0$ and $\omega_c$. This is the primary degeneracy between the PPN and $\Lambda$CDM parameters.
   \item Our results are complementary to existing PPN constraints, as they come from different observables, with different assumptions and modelling, and from different length and time scales. They ensure that cosmological modelling is automatically consistent with Solar System tests, while nonetheless constraining a single unified set of parameters.
\end{itemize}

These results comprise an initial study into constraining the PPN parameters with cosmological data, and combining them with data on other scales. There is significant scope to improve and extend the results of this study, notably by relaxing some of the additional freedom that can be allowed in the cosmology, such as allowing a non-constant behaviour for $\gamma_c$ or allowing the large-scale slip to differ from its GR value, or by including additional cosmological data sets such as those from Baryon Acoustic Oscillations. One could also directly include Solar System and astrophysical constraints as priors on $\gamma$ and $\dot{\alpha}$ at the present time, which would automatically ensure that cosmological modelling is consistent with observations on these scales without invoking any hypothetical screening mechanisms. Theoretical work extending the Effective Field Theory of Large Scale Structure (see e.g. \cite{Baumann:2010tm}) to the PPNC framework, or studies of non-linear structure formation in the PPNC framework, would further broaden the range of observables that can be combined for precision tests of gravity. The assumption of power-law evolution could also be dropped in favour of something more directly reconstructed from the data, such as Gaussian Processes or Genetic Algorithms, which we expect would further show the advantages of including cosmological observables in the same framework as astrophysical and Solar System tests.

In summary, the PPNC framework demonstrated here increases the robustness and scope of non-cosmological tests of gravity, and allows the time evolution of the PPN parameters to be constrained in ways that were previously impossible. These results pave the way for unified precision tests of GR combining cosmological, astrophysical and Solar System constraints.

\section*{Acknowledgements}
We gratefully acknowledge useful discussions with Mike Wilensky and Chris Addis. DBT, TA and TC acknowledge support from the Science and Technology Facilities Council (STFC, grant numbers ST/P000592/1 and ST/X006344/1). This result is part of a project that has received funding from the European Research Council (ERC) under the European Union's Horizon 2020 research and innovation programme (Grant agreement No. 948764; PB). This research utilised Queen Mary's Apocrita HPC facility, supported by QMUL Research-IT \href{http://doi.org/10.5281/zenodo.438045}{http://doi.org/10.5281/zenodo.438045}. We acknowledge the assistance of the ITS Research team at Queen Mary University of London. We acknowledge the use of GetDist \cite{getdist}.

\appendix 

\section{Code Implementation}
\label{app_code}

Here we present the form of the equations described in Section \ref{theory}, using the conventions of the CLASS code for perturbations.

\subsection{Implemented equations}

The modified Friedmann equation for the background evolution is
\begin{eqnarray}
&& H=\sqrt{\frac{8\pi G}{3}\left(\bar{\rho}_\text{tot}-\bar{\rho}_c-\bar{\rho}_b\right)+\frac {8\pi G}{3}\left(\bar{\rho}_c+\bar{\rho}_b\right)\gamma-\frac{2\gamma_c}{3}-\frac{K}{a^2}} \,\text{,}
\end{eqnarray}
where the present-day value of $\gamma_c$ is calculated from the Friedmann equation at $a=1$,
\begin{eqnarray}
\gamma_{c,0}&=&\frac{3}{2}\Omega_m H^2_0 (\gamma_0-1)+\gamma_{c,\Lambda}\,\text{.}
\end{eqnarray}
A subscript $0$ denotes a quantity evaluated at the present time $a=1$, and $\gamma_{c,\Lambda}$ is the value of $\Lambda$ that would be required in a flat universe given the other $\Lambda$CDM parameter values. The parameter $\alpha_c$ is then obtained from the integrability condition (\ref{eq_integrability}) as
\begin{eqnarray}
\alpha_c&=&\frac{3}{2}\Omega_m H^2_0 a^{-3}\left( \alpha-\gamma+\hat{\gamma}\right)-2\gamma_c-\hat{\gamma}_c\,\text{,}
\end{eqnarray}
where hats denote differentiation with respect to $\ln a$.

The equations of motion are integrated forward in time, with the PPNC parameters taking non-GR values from an initial scale factor $a_1=10^{-10}\,$ (although the exact value does not matter, as long as it is deep in the radiation era). For the power law behaviour considered in this work, once the averaged PPN parameters $\bar{\alpha}$ and $\bar{\gamma}$ from Eq. (\ref{agbar}) are specified, the initial values of $\alpha$ and $\gamma$ at $a = a_1$ can be reconstructed from the average values using
\begin{eqnarray}
    \alpha(a_1)&=&\left[1-\frac{1}{1-a^n_1}-\frac{1}{n\ln(a_1)}\right]^{-1}\left(\bar{\alpha}-\frac{\alpha_0}{1-a^n_1}-\frac{\alpha_0}{n(\ln(a_1))}\right) \, ,
\end{eqnarray}
with a similar expression for $\gamma(a_1)$. In general, the code is designed to allow for both $\gamma_c$ and the large-scale slip to vary in time with power-law behaviour; however we do not use these options for this work. We note that for the theories considered in Ref. \cite{Thomas_2023} a single power-law over all times is not a good description of the behaviour of $\gamma_c$\,. However it is a reasonable approximate description at late times, and the contribution at early times is small. 

The scalar perturbations to the metric are governed by
\begin{eqnarray}
\psi&=&\eta \phi - \frac{12\pi G a^2}{k^2} \left[\left(\bar{\rho}+\bar{p}\right)\sigma\right]_\text{tot}\,, \\
\phi_{,\tau} &=& -\mathcal{H}\psi+\frac{4\pi G a^2}{k^2} \left(\left[(\bar{\rho}+\bar{p})\theta\right]_\text{tot}-\tilde{A}_m \right)+\frac{4\pi G a^2}{k^2} \mu \tilde{A}_m +\mathcal{G}\mathcal{H}\phi\,,
\end{eqnarray}
where $\tilde{A}_m =\bar{\rho}_b \theta_b+ \bar{\rho}_c \theta_c $. The code potentials are related to those used in Section \ref{theory} by $\psi = -\Phi$ and $\phi = -\Psi\,$, and $\theta = - v \, k^2$. The PPNC formalism is most naturally defined in the Newtonian gauge \cite{Clifton:2020oqx}, so this choice is used throughout the modified code, and there is no synchronous gauge implementation. The interpolations (at any given time) for a general wavenumber $k$ between the large ($L$) and small-scale ($S$) values of $\mu$, $\eta$ and $\mathcal{G}$ are given by \cite{Thomas_2023}
 \begin{equation}
   f(k,a)=\frac{1}{2}(S+L)+\frac{1}{2}(S-L)\tanh\left(\log\left[k/\left(\frac{a_{,\tau}}{a}\right)\right]\right) \text{.}
   \end{equation}
For the purpose of the studies carried out in this work, the code allows users to evolve the perturbations without the additional $\mathcal{G}$ term in the momentum constraint equation by effectively setting $\mathcal{G} = 0$ always. The code can be run with either background or perturbation equations modified, or both. The code will be made public in future with full documentation of the modifications.


\section{Additional plots and tables}
\label{app_constraints}

Here we will present some additional results that were not shown in the main body of the paper in order to keep the presentation as concise as possible.

\begin{figure}
    \hspace{-2cm}
    \includegraphics[width=0.57\linewidth]{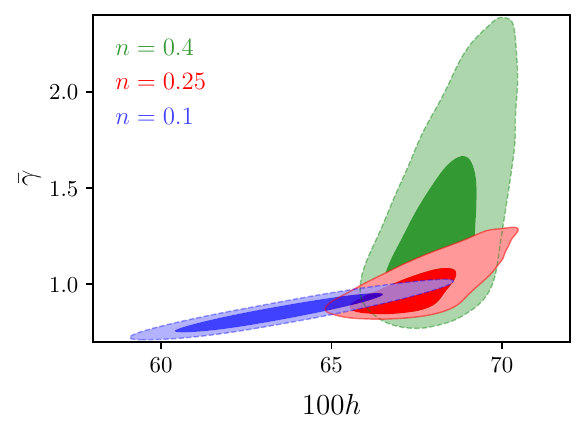}
    \includegraphics[width=0.57\linewidth]{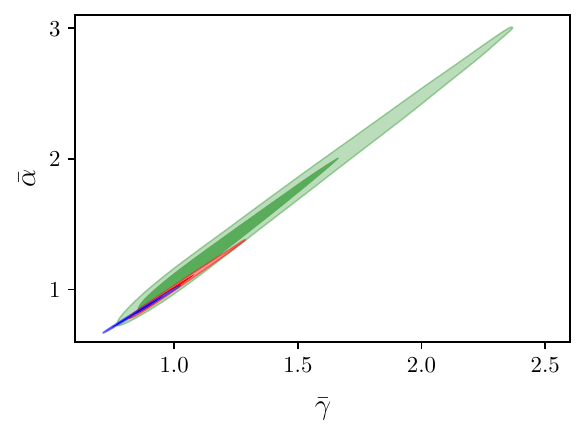}
    \caption{2D posteriors for $\left(h, \bar{\gamma}\right)$ and $\left(\bar{\gamma},\bar{\alpha}\right)\,$\,, for the fixed power laws with $n = 0.1$\,, $0.25$ and $0.4\,$. There is a strong increase in the allowed range of PPN parameters as the power-law index is increased.}
    \label{fig_h0_alphagamma_2d_2}
\end{figure}

Fig. \ref{fig_h0_alphagamma_2d_2} shows the effect on the degeneracies between $\bar{\gamma}$ and $h$\,, and between $\bar{\alpha}$ and $\bar{\gamma}$\,, of increasing the power law into the physically uninteresting region $n \gtrsim 0.25\,$. One sees clearly that the $\left(h, \bar{\gamma}\right)$ degeneracy becomes much weaker as the power law enters this region, as the 1D confidence interval on $\bar{\gamma}$ widens drastically. The weakening of the $\left(\bar{\gamma}, \bar{\alpha}\right)$ degeneracy is somewhat less drastic.

\begin{figure}
    \centering
    \begin{subfigure}{0.85\textwidth}
    \includegraphics[width=\linewidth]{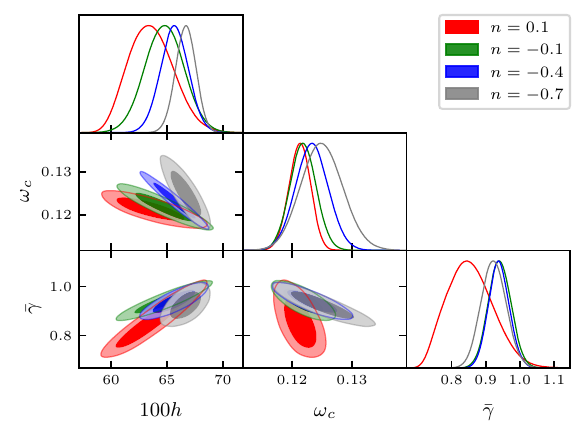}
    \end{subfigure}
    \begin{subfigure}{0.85\textwidth}
    \includegraphics[width=\linewidth]{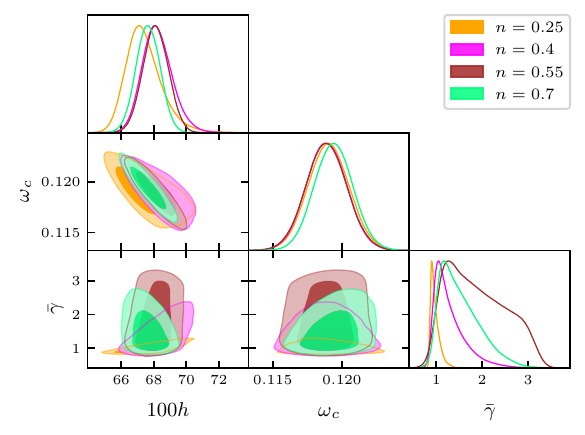}
    \end{subfigure}
    \caption{2D and 1D posteriors on $h\,$, $\omega_c$\,, and $\bar{\gamma}$\,, for the $n = 0.1$\,, $-0.1$\,, $-0.4$ and $-0.7$ power laws (above), and the $n = 0.25\,$, $0.4$\,, $0.55$ and $0.7$ power laws (below). For the lower panel, a smoothing scale of $0.3\sigma$ has been applied to the posteriors.}
    \label{fig_H0_omc_gamma_corner_both}
\end{figure}

In Fig. \ref{fig_H0_omc_gamma_corner_both}, we display the full set of 1D and 2D posteriors on $\bar{\gamma}\,$, $h$ and $\omega_c\,$, for PPNC power laws in the range $-0.7 \leq n \leq 0.7\,$. The $n = 0.1$ power law (red contours in the upper panel) is particularly notable, as the 1D posterior distribution on $\bar{\gamma}$ is broader than other power laws in the physically interesting region, and has its mean further from unity, with corresponding implications for the posteriors on $h$ and $\omega_c\,$.

Tables \ref{table_all_powerlaws_summary} and \ref{table_derivatives_and_n} respectively show the $95\%$ confidence interval constraints on the standard $\Lambda$CDM cosmological parameters and the time-weighted means of the PPN parameters, and the present-day derivatives of the PPN parameters with respect to the scale factor (equivalently, their time derivatives at the present day, divided by $H_0$). Results are shown for PPNC power laws between $n = -1$ and $n = 0.4\,$, and for the varying power law. 

\begin{figure} \nonumber
    \centering
    \begin{subfigure}{0.75\textwidth}
    \includegraphics[width=\linewidth]{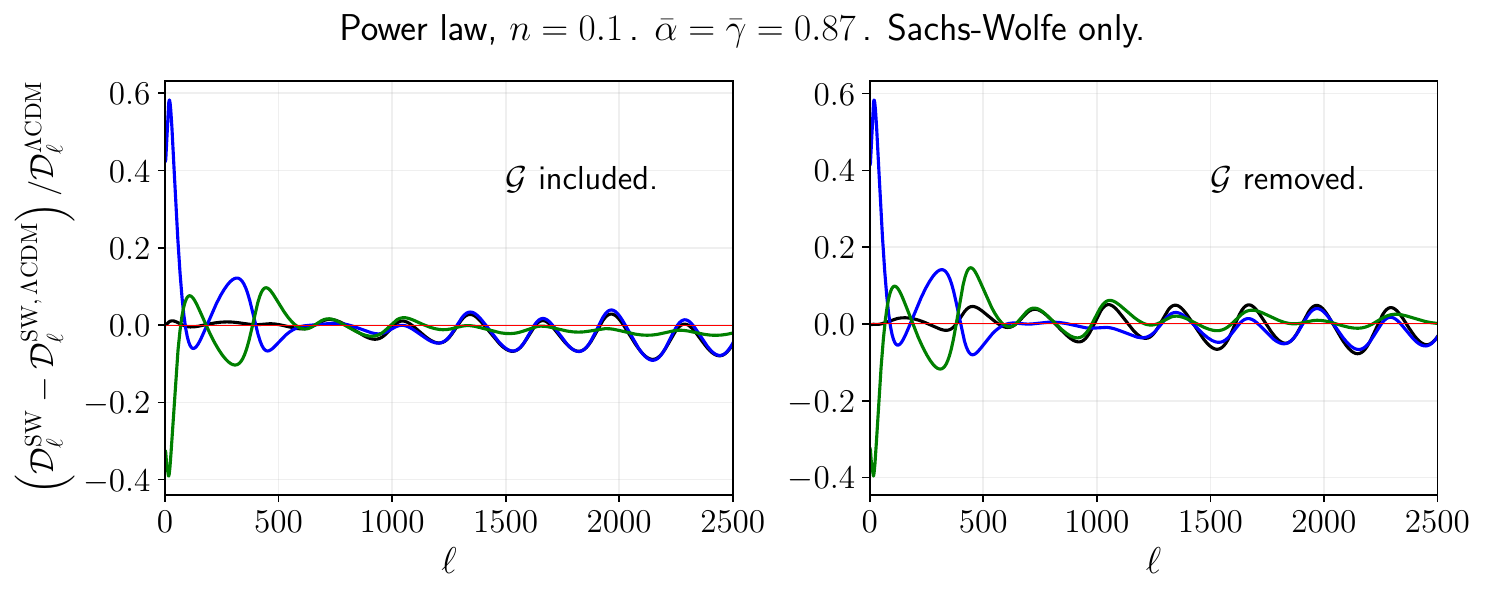}
    \end{subfigure}
    \begin{subfigure}{0.75\textwidth}
    \includegraphics[width=\linewidth]{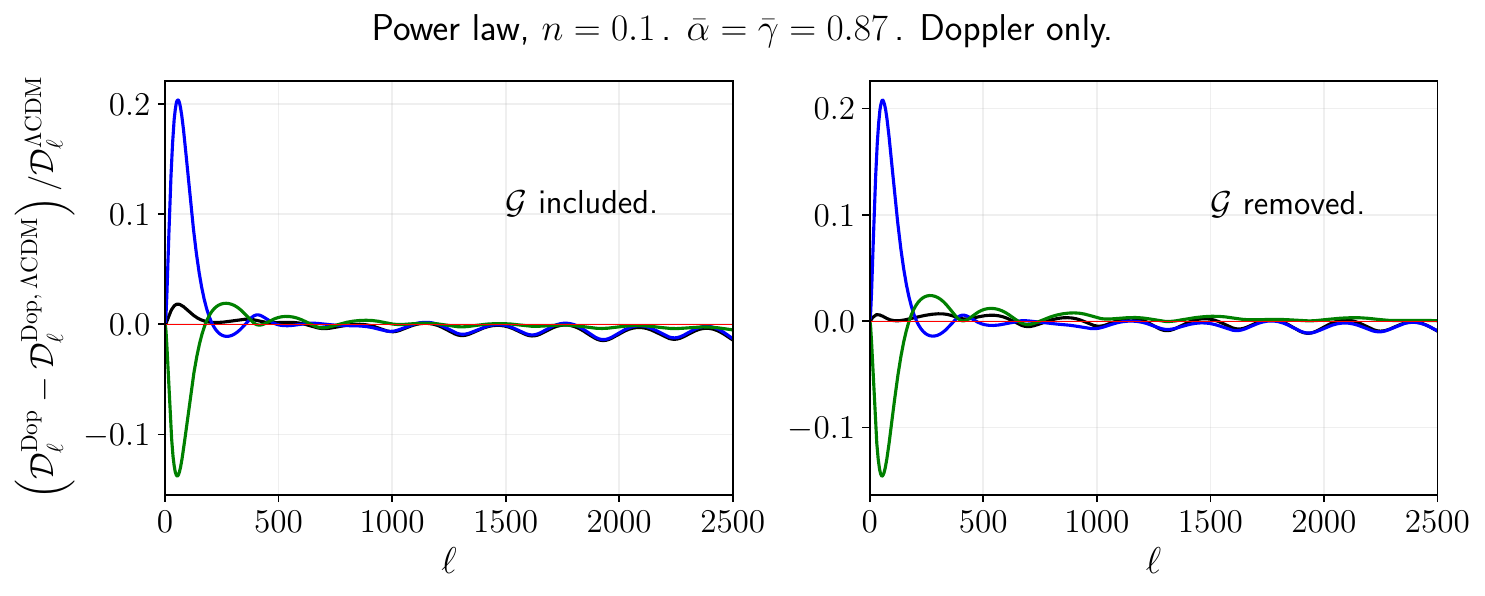}
    \end{subfigure}
    \begin{subfigure}{0.75\textwidth}   \includegraphics[width=\linewidth]{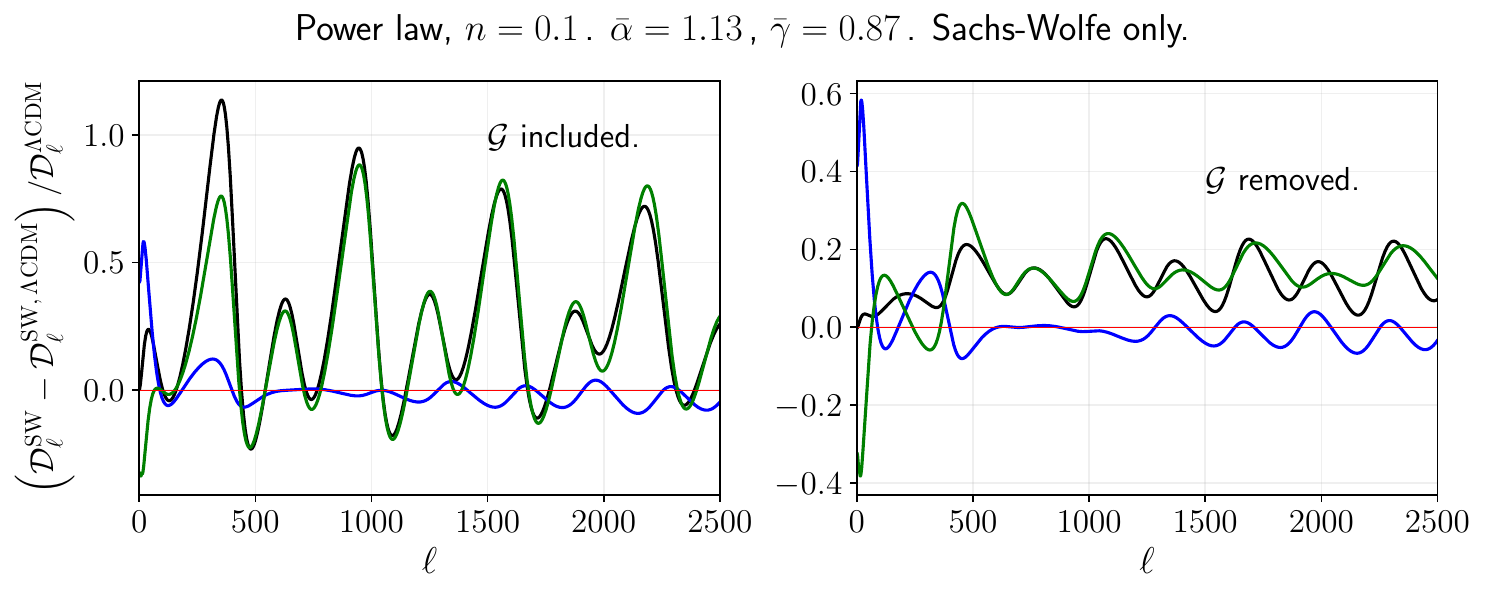}
    \end{subfigure}
    \begin{subfigure}{0.75\textwidth}
    \includegraphics[width=\linewidth]{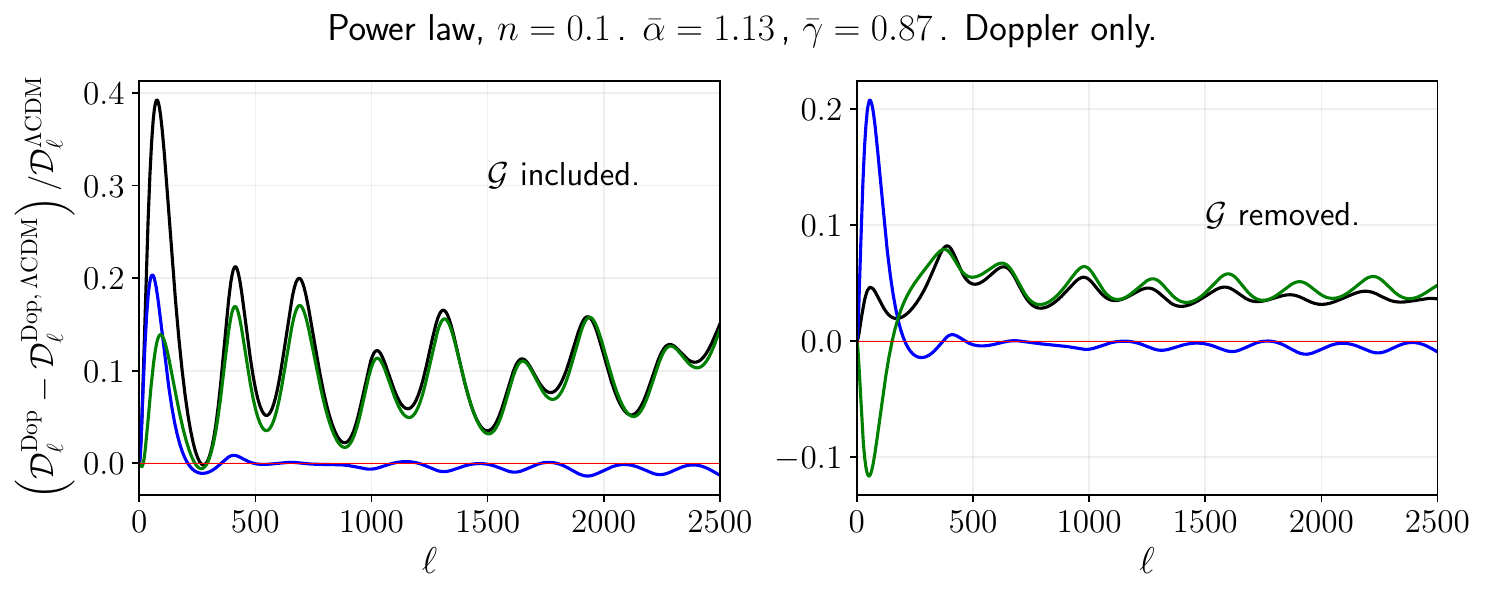}
    \end{subfigure}
    \caption{The relative effects of a PPNC background evolution and PPNC perturbation evolution on the Sachs-Wolfe (first and third) and Doppler (second and fourth) terms. In each case, we have first used the full set of PPNC equations, and then re-calculated with the $\mathcal{G}$ PPNC function artificially removed, in order to demonstrate its importance. Results are shown for $n = 0.1$. The top two rows have $\bar{\alpha} = \bar{\gamma} = 0.87\,$, and the bottom two rows have $\bar{\alpha} = 1.13$ and $\bar{\gamma} = 0.87\,$. Black curves show the results of a full PPNC evolution, green shows when the PPNC perturbation equations are used but the background expansion is $\Lambda$CDM, blue shows when the background is evolved using the PPNC Friedmann equation but the perturbations are evolved according to $\Lambda$CDM equations, and the flat red lines indicate $\Lambda$CDM. The plots show similar behaviour to Fig. \ref{fig_background_vs_perturbations}.}
\label{fig_background_vs_perturbations_2}
\end{figure}

In Fig. \ref{fig_background_vs_perturbations_2}, we show the equivalents to Fig. \ref{fig_background_vs_perturbations}, but rather than evaluating the residuals relative to $\Lambda$CDM in the full angular power spectrum $\mathcal{D}_{\ell}\,$, we focus on the residuals in the Sachs-Wolfe and Doppler source contributions specifically, as it is these terms that primarily drive the requirement that $\bar{\alpha} \approx \bar{\gamma}\,$.

\begin{table}
\hspace{-1cm}
\begin{mytabular}[1.2]{|l|c|c|c|c|c|c|c|c|c|c|} 
\hline 
Power law & $-1$ &$-0.7$ & $-0.4$ & $-0.1$ & $0.1$ & $0.25$ & $0.4$ & varying \\ 
\hline 
$100\,\omega_b$ & $2.24_{-0.05}^{+0.05}$ & $2.24_{-0.05}^{+0.05}$ & $2.24_{-0.05}^{+0.05}$ & $2.24_{-0.05}^{+0.05}$ & $2.23_{-0.05}^{+0.05}$ & $2.21_{-0.05}^{+0.05}$ & $2.21_{-0.05}^{+0.05}$ & $2.23_{-0.06}^{+0.05}$ \\ 
\hline
$10\,\omega_c$ & $1.27_{-0.09}^{+0.09}$ & $1.25_{-0.07}^{+0.07}$ & $1.23_{-0.06}^{+0.06}$ & $1.22_{-0.05}^{+0.05}$ & $1.21_{-0.04}^{+0.04}$ & $1.19_{-0.03}^{+0.03}$ & $1.19_{-0.03}^{+0.03}$ & $1.21_{-0.04}^{+0.04}$ \\ 
\hline
$\ln{10^{10}\,A_s}$ & $3.02_{-0.04}^{+0.04}$ & $3.02_{-0.04}^{+0.04}$ & $3.02_{-0.04}^{+0.04}$ & $3.03_{-0.03}^{+0.03}$ & $3.04_{-0.03}^{+0.03}$ & $3.04_{-0.03}^{+0.03}$ & $3.03_{-0.03}^{+0.03}$ & $3.03_{-0.03}^{+0.03}$ \\ 
\hline
$10\, n_s$ & $9.72_{-0.18}^{+0.18}$ & $9.71_{-0.17}^{+0.17}$ & $9.69_{-0.17}^{+0.17}$ & $9.67_{-0.16}^{+0.16}$ & $9.69_{-0.16}^{+0.16}$ & $9.62_{-0.15}^{+0.15}$ & $9.59_{-0.15}^{+0.15}$ & $9.66_{-0.16}^{+0.16}$\\ 
\hline
$100\,\tau$ & $5.02_{-1.72}^{+1.71}$ & $4.94_{-1.75}^{+1.74}$ & $4.88_{-1.73}^{+1.71}$ & $5.00_{-1.65}^{+1.72}$ & $5.00_{-1.66}^{+1.65}$ & $5.41_{-1.56}^{+1.62}$ & $5.47_{-1.54}^{+1.59}$ &  $5.13_{-1.65}^{+1.62}$ \\ 
\hline
$10\,Y_p$ & $2.32_{-0.32}^{+0.31}$ & $2.33_{-0.32}^{+0.32}$ & $2.34_{-0.32}^{+0.32}$ & $2.35_{-0.34}^{+0.34}$ & $2.32_{-0.34}^{+0.34}$ & $2.23_{-0.35}^{+0.35}$ & $2.21_{-0.36}^{+0.36}$ & $2.30_{-0.35}^{+0.35}$ \\ 
\hline
$100\, h$ & $67.6_{-1.6}^{+1.6}$ & $66.7_{-1.9}^{+1.9}$ & $65.7_{-2.5}^{+2.6}$ & $64.8_{-3.6}^{+3.6}$ & $63.7_{-3.9}^{+4.0}$ & $67.4_{-2.2}^{+2.5}$ & $68.2_{-1.9}^{+2.0}$ &  $65.1_{-4.1}^{+3.9}$ \\ 
\hline
$\bar{\gamma}$ & $0.91_{-0.09}^{+0.09}$ & $0.92_{-0.07}^{+0.08}$ & $0.94_{-0.06}^{+0.06}$ & $0.94_{-0.07}^{+0.07}$ & $0.86_{-0.13}^{+0.13}$ & $0.97_{-0.14}^{+0.22}$ & $1.29_{-0.40}^{+0.75}$ &  $0.90_{-0.15}^{+0.14}$ \\ 
\hline
$\bar{\alpha}$ & $0.91_{-0.09}^{+0.09}$ & $0.92_{-0.07}^{+0.08}$ & $0.94_{-0.06}^{+0.06}$ & $0.94_{-0.07}^{+0.07}$ & $0.84_{-0.16}^{+0.15}$ & $0.98_{-0.15}^{+0.28}$ & $1.47_{-0.58}^{+1.07}$ & $0.89_{-0.18}^{+0.17}$ \\ 
\hline
\end{mytabular}
\caption{Constraints for fixed power-laws with $n \in \left\lbrace -1, -0.7, -0.4, -0.1, 0.1, 0.25, 0.4\right\rbrace$\,, and for a varying power law with a flat prior on $n$ between $-15$ and $0.25\,$. For all parameters, we show their mean and $95\%$ confidence intervals\,.}
\label{table_all_powerlaws_summary}
\end{table}

\begin{table}
\centering
\begin{mytabular}[1.2]{|l|c|c|} 
\hline 
Power law & $\dot{\gamma}_0/H_0$ & $\dot{\alpha}_0/H_0$ \\ 
\hline 
$-1$ & $0.098_{-0.096}^{+0.092}$ & $0.096_{-0.089}^{+0.091}$  \\
\hline
$-0.7$ & $0.057_{-0.057}^{+0.056}$ & $0.056_{-0.057}^{+0.056}$  \\
\hline
$-0.4$ & $0.027_{-0.028}^{+0.028}$ & $0.027_{-0.029}^{+0.028}$  \\
\hline
$-0.1$ & $0.001_{-0.011}^{+0.011}$ & $0.010_{-0.012}^{+0.012}$ \\
\hline
$0.1$ & $0.0049_{-0.0045}^{+0.0045}$ & $0.0056_{-0.0051}^{+0.0054}$  \\
\hline
$0.25$ & $\left(1.3_{-1.0}^{+6.3}\right)\times 10^{-4}$ & $\left(1.1_{-1.3}^{+6.9}\right) \times 10^{-4}$  \\
\hline
$0.4$ & $\left(-1.1_{-2.8}^{+1.5}\right) \times 10^{-4}$ & $\left(-1.7_{-3.9}^{+2.1}\right)\times 10^{-4}$ \\
\hline
varying & $0.0047_{-0.0080}^{+0.0141}$ & $0.0051_{-0.0095}^{+0.0134}$\\ 
\hline
\end{mytabular}
\caption{$95\%$ constraints on the present-day time derivatives of the post-Newtonian parameters as a ratio of $H_0$\,, for the fixed PPNC power laws with power-law index $n \in \left\lbrace -1, -0.7, -0.4, -0.1, 0.1, 0.25, 0.4\right\rbrace$\,, and for a varying power-law with a flat prior on $n$ between $-15$ and $0.25\,$.}
\label{table_derivatives_and_n}
\end{table}

\bibliographystyle{unsrt}
\bibliography{ppnc}

\end{document}